\documentclass[linenumbers]{aa}  

\usepackage{color}
\usepackage{chemformula}

\newcommand{\lalpha}{Ly$\alpha$}
\newcommand{\halpha}{H$\alpha$}
\newcommand{\hi} {H{\sc i}} 
\newcommand{\hp}{H$^+$}
\newcommand{\hei} {He {\sc i}}

\usepackage{graphicx}
\usepackage{txfonts}
\usepackage{amsmath}
\usepackage{multirow}
\usepackage{ulem}

\usepackage{natbib,twoopt}
\usepackage[breaklinks=true]{hyperref} 
\bibpunct{(}{)}{;}{a}{}{,}             

\begin{document}

   \title{
   Self-consistent simulation of photoelectrons in exoplanet winds: Faster ionisation and weaker mass loss rates.}

   \author{A. Gillet 
          \inst{1}
          \and
          A. Garc\'ia Mu\~noz\inst{1}
          \and
          A. Strugarek\inst{1}
          }

   \institute{Université Paris Cité, Université Paris-Saclay, CEA, CNRS, AIM, F-91191, Gif-sur-Yvette, France\\
              \email{alexandre.gillet@cea.fr}
             }


  \abstract
   {Close-in exoplanets undergo extreme irradiation levels leading to hydrodynamic atmospheric escape and the formation of planetary winds. The planetary mass loss is governed by several physical mechanisms including photoionisation that may impact the evolution of 
   the atmosphere. The stellar radiation energy deposited as heat depends strongly on the energy of the primary electrons following photoionisation and on the local fractional ionisation. 
   All these factors affect the model-estimated atmospheric mass loss rates and other characteristics of the outflow in ways that have not been clearly elucidated.
   Moreover, the shape of the XUV stellar spectra influences strongly the photoionisation and heating deposition on the atmosphere. Substantial changes are to be expected on the planetary mass loss rate.  
 }
{We study the effect of secondary ionisation by photoelectrons on the ionisation and heating of the gas for different planet-star systems. We elaborate on the local and planet-wise effects, to clearly demonstrate the significance of such interactions.
}
   {
   Using the PLUTO code, we performed 1D hydrodynamics simulations for a variety of planets and stellar types. We include planets in the range from Neptune to Jupiter size, and stars from M dwarfs to Sun-like.
   }
   {Our results indicate a significant decrease of the planetary mass loss rate for all planetary systems when secondary ionisation is taken into account. The mass loss rate is found to decrease by 43$\%$ for the more massive exoplanet to 54$\%$ for the less massive exoplanet orbiting solar-like stars, and up to 52$\%$ for a Jupiter-like planet orbiting a M type star.
   Our results also indicate much faster ionisation of the atmosphere due to photoelectrons.
   }
  {We built a self-consistent model including secondary ionisation by photoelectron to evaluate its impact on mass loss rates. We find that photoelectrons affect the mass loss rates by factors that are potentially important for planetary evolution theories. We also find that enhanced ionisation occurs at altitudes that are often probed with specific atomic lines in transmission spectroscopy. Future modeling of these processes should include the role of photoelectrons. To that end, we make available a simple yet accurate parameterisation for atomic hydrogen atmospheres.
  }


   \maketitle
%

\section{Introduction} \label{sec:intro}

It has long been proposed that the atmospheres of exoplanets orbiting close-in 
to their host stars must be rapidly escaping \citep{lammer2003atmospheric,baraffe2004effect}. Accurately predicting the mass loss rates remains however a difficult task due to the complexity of the star-planet interactions that participate in the mass loss process. 
\par
Planets can lose their atmospheres through multiple mechanisms, broadly separated into thermal and non-thermal \citep[see][for a complete review on escape processes]{gronoff2020atmospheric}.
Non-thermal processes such as ion sputtering, polar outflow or charge-exchange with the stellar wind dominate for planets orbiting at large distances from their star. In the context of planets on short-period orbits, thermal processes, which include Jeans and hydrodynamic escape, generally dominate.
We will focus here on hydrodynamic escape, which occurs when intense stellar radiation deposits its energy in the planetary atmosphere (at various altitudes depending on the radiation wavelength), leading to the escape of the atmosphere into space.
\par
The planetary wind interacts with the stellar wind and forms a number of hydrodynamic features such as a bow shock, a comet-like tail and Kelvin–Helmholtz instabilities \citep{tremblin2013colliding}. Insight into these processes can be gained by measuring with in-transit spectroscopy the excess absorption that occurs at a number of atomic lines such as {\hi} {\lalpha} and {\halpha}, or the {\hei} triplet at 10830 {\AA}.
These features have been measured for hot Jupiters \citep[e.g.][]{vidal2003extended,2010A&A...514A..72L,jensen2012detection} and hot Neptunes \citep[e.g.][]{kulow2014lyalpha,ehrenreich2015giant,ben2022signatures}. A number of studies have been conducted to understand the global problem of  planetary escape with hydrodynamical 1D models \citep[e.g.][]{yelle2004aeronomy,garciamunoz2007physical,murray2009atmospheric,koskinen2013escape}.
Such models predict velocities for the planetary wind of a few km/s in the planet's proximity, which are consistent with the line broadening of {\halpha} and the {\hei} triplet measured with high-resolution spectroscopy \citep[e.g.][]{salz2018detection,munoz2019rapid}. 
However, they fail to predict the velocities of $\sim$100 km/s that are found with {\lalpha} spectroscopy \citep[e.g.][]{vidal2003extended,ben2007exoplanet}. Whether the velocities measured with {\lalpha} spectroscopy are representative of the planetary wind or are instead indicative of stellar wind protons that become neutralised by charge-exchange with the planetary wind remains an open question \citep{tremblin2013colliding}.

Many authors have looked into this problem with 2D and 3D models that include self-consistent radiative transfer, with the goal of constraining the mass loss rates from the planets \citep[e.g.][]{tripathi2015simulated,debrecht2019photoevaporative}. Recently, \citet{shaikhislamov2021global} developed a global 3D multi-fluid model to investigate the He 10830 {\AA} line; \citet{daley2019hot} explored the star-planet interaction of the wind in the presence of a magnetic field; and \citet{carolan2021effects} looked at the effects of the stellar wind strength on a magnetised planetary wind, showing that the increased polar loss compensates for the decreased mass loss at the dead zones of the planet.
\par
Accurately predicting the atmospheric mass loss rate is essential to answering fundamental questions about the evolution of planets. Indeed, some atmospheres, in particular those of low-mass planets, may be escaping so efficiently that they can be completely lost to space on timescales shorter than the planet's lifetime. In the case of hydrodynamic escape of strongly irradiated exoplanets, the aim of this work, calculating the mass loss rate involves determining how much of the stellar radiation energy is converted into heat, and how much goes otherwise to ionisation and excitation of the atmospheric gas. 
\par
We focus here on atmospheric gas made of hydrogen atoms H because although H$_2$ should be prevalent in primary atmospheres, the molecule will typically dissociate rapidly in the planet's upper atmosphere. Also, the physics of photoelectron interactions with H$_2$ is more complex than for H atoms because, 
being a molecule, H$_2$ offers additional channels for excitation, dissociation and ionisation 
that must be tracked individually \citep{hallettetal2005}. We postpone such a study to future work and focus here on the essentials for an atomic atmosphere. Upon absorption by the atmospheric atomic gas, the stellar X-ray and Extreme Ultraviolet photons (jointly referred to as XUV, and covering wavelengths from a few {\AA} to the Lyman continuum threshold at 912 {\AA}) release high velocity electrons.
These so-called photoelectrons have energies $E_{\rm{0}}$=$hc$/$\lambda${$-$13.6 eV}, where $h$ and $c$ are Planck's constant and the speed of light, respectively, and $\lambda$ is the photon wavelength. The photoelectrons can excite the gas and produce secondary electrons while slowing down. The gas heating corresponds to the fraction of the photoelectron's initial energy ${E_0}$ that is not expended in excitation or ionisation but that goes instead into kinetic energy.
\par
We emphasize the importance of two parameters that dictate the fraction of the photoelectron's initial energy that is deposited as heat, namely: the local fractional ionisation $x_e$ and the energy $E_0$ of the primary photoelectrons. 
In particular, if the local fractional ionisation is high, elastic collisions between the fast and thermal electrons ensure that most of $E_0$ is transferred to the thermal electrons, thereby heating the gas. Also, if $E_0$ is less than 10.2 eV, i.e. below the lowermost threshold for inelastic collisions in the H atom, 
all of $E_0$ is transferred as heat to the background gas in elastic collisions. In these two limits, the prediction of the gas heating is relatively straightforward. Importantly, when the fractional ionisation is low or moderate and $E_0$ is sufficiently large, additional electrons and ions are created during the secondary ionisation process, when the primary photoelectron interacts with other hydrogen atoms. For the most energetic primary electrons created after photoionisation, it follows a cascade of multiple ionisation events.
\par
The fundamental information for the treatment of photoelectrons, and for the assessment of how much of their energy goes into excitation, ionisation and heating of the gas has been available for decades. 
For example, \citet{habing1971heating} and \citet{shull1979heating} have looked at the production of secondary electrons induced by soft X-rays and how they interact with a primordial gas using a Monte Carlo method. The outcome of such modelling efforts can be readily parameterised as a function of $x_e$ and $E_0$ to take into account the effect of photoelectrons in the net ionisation rate and heating rate induced by XUV photons impacting the upper atmosphere of short-period orbit exoplanets. 
 
\par
In this work, we revisit the problem by exploring self-consistently the simultaneous heating
and ionisation that occurs as the photoelectrons slow down, taking advantage of recent advances in the characterisation of the XUV spectra of stars. This effect is often neglected in the studies dedicated to modelling hydrodynamic escape of planetary atmospheres 
\citep[e.g.][]{lammer2003atmospheric,wu2013density,munoz2020pi,munoz2021heavy}. With the ultimate goal of better predicting the loss rate and the main features of planetary winds, in some cases an arbitrary pre-fixed fraction (typically, 2-30{\%}) of the photoelectron energy $E_0$ is assumed to transfer into heating \citep{linssen2022constraining}. Our work shows how to partly overcome such simplified treatments and elaborates on the importance of photoelectrons in the strongly-irradiated atmospheres of some exoplanets.
\par
\citet{guo2016influence} have covered related ideas, with an emphasis on how the spectral energy distribution of the star affects the mass loss rate and ionisation of strongly-irradiated atmospheres. 
Following their initial study, in this work we put significant effort to elucidate how the photoelectrons affect the atmosphere both locally and globally, and provide simple yet accurate descriptions for the self-consistent implementation of such processes in the continuity and energy conservation equations of the gas. These prescriptions should be useful to other modelling efforts. In addition, we leverage recent developments in the reconstruction of the XUV spectra for cool stars by observations at X-ray and far ultraviolet wavelengths \citep{france2016muscles}. 
\par
To examine the quantitative effect of photoelectrons and assess whether the planet's gravity plays a role, we performed 1D spherical hydrodynamic simulations with the PLUTO code \citep{mignone2007pluto}. We focused on the escaping atmospheres of four hypothesised planetary systems with masses ranging from 0.02 $M_{J}$ to 0.69 $M_{J}$ (similar to HD209458b) and assessed the effect of different XUV spectra  taken from the MUSCLES survey \citep{france2016muscles}. 
We defined the planet size $R_p$ so that the planet density remains constant in all cases. This is a somewhat arbitrary choice, but useful because of its simplicity. In the energy-limited limit, the mass loss rate \citep{erkaev2007roche} is proportional to the inverse of the planet density, in which case assuming a constant bulk density serves as a valid basis to compare against that limit. Our simulations incorporate the relevant physics of photoelectrons in the continuity and energy equations.
\par
The plan of the paper is the following: the model, including the PLUTO setup and our treatment of radiative transfer with photoelectrons, is described in section \S \ref{sec:modeldescription}. 
In section \S \ref{sec:secondary}, we describe the relevant physics of photoelectrons and its implementation in PLUTO. Section  \S \ref{sec:outflowdescription} is dedicated to the description of the results of the atmospheric escape of a Neptune-like planet. Section  \S \ref{sec:massdependency} describes the dependence of our results on the planet mass, and finally in section \S \ref{sec:stellarspectra} we study the impact of the shape of the stellar spectra of M and K type stars on secondary ionisation.

\section{Model description} 
\label{sec:modeldescription}
\subsection{Physical model}
The model is constructed with the hydrodynamics code PLUTO \citep{mignone2007pluto}, which solves the Euler equations in a rotating reference frame. The 1D equations for conservation of mass, momentum and energy solved in PLUTO are:

\begin{equation}
\label{eq:1}
\frac{ \partial \rho}{ \partial t} + \boldsymbol{\nabla} \cdot (\rho {\bf u})= 0 \, ,
\end{equation}

\begin{equation}
\label{eq:2}
\frac{ \partial (\rho{\bf u})}{ \partial t} + \boldsymbol{\nabla} \cdot \left[\rho {\bf u} {\bf u} + P_T {\bf I}  \right] = - \rho \nabla \phi + \rho\bf{F}_{\rm cent}\, ,
\end{equation}

\begin{equation}
\label{eq:3}
\frac{ \partial (E + \rho\phi)}{ \partial t} + \boldsymbol{\nabla} \cdot [(E + P_T + \rho\phi){\bf u}] = \rho {\bf F}_{\rm cent}{\bf u} + H - C \, ,
\end{equation}
where $r$ is distance as measured from the planet's centre and $t$ is time, and where ${\bf u}$ is the fluid velocity, $\rho$ the density, $P_T$ the thermal pressure, $E = P_T /(\gamma -1) + \rho {\bf u}^2/2$ the total energy with $\gamma$ = 5/3, the adiabatic index for a mono-atomic gas, and $C$ and $H$ are the cooling and heating terms to be defined in \S \ref{sec:HeatingAndCooling}. The joint gravitational potential of the planet plus the star is defined as:
\begin{equation}
    \label{eq:GravPot}
    \phi = -\frac{GM_p}{r} - \frac{GM_\star}{R_{\rm orbit}-r}\, ,
\end{equation}
where $G$ is the gravitational constant, and $M_\star$ and $M_p$ are the mass of the star and the planet, respectively.
Finally, the centrifugal force is written as:
\begin{equation}
    \label{eq:Fcent}
    {\bf F}_{\rm cent} = - \frac{GM_\star}{R_{\rm orbit}^3}\left( R_{\rm orbit} - r \right) {\bf e}_r\,,
\end{equation}
 where $R_{\rm orbit}$ is the distance from the planet centre to the stellar centre. ${\bf e}_r\,$ is the unit vector in the outwards radial direction.
\par
We consider here an atmosphere composed of atomic hydrogen in neutral {\hi} and ionised H$^{+}$ forms, plus thermal electrons. The density in the equations above corresponds to the total mass density $\rho = \rho_{\text{\hi}} + \rho_{\text{\hp}}$, which neglects the contribution of thermal electrons. We define the neutral fraction $x_{\text{\hi}}=\rho_{\text{\hi}}/\rho$ and the fractional ionisation as $x_e=1-x_{\text{\hi}}$, both taking values between 0 and 1. On the basis of charge neutrality, the number density of ions and electrons is the same. To determine the partitioning between {\hi}, {\hp} and electrons, we consider the processes for:
\par
collisional ionisation: \ch{H + e- -> \text{\hp} + 2 e-}\,,
\par
radiative recombination: \ch{\text{\hp} + e- -> H + h$\nu$}\,,
\par
photoionisation: \ch{H + h$\nu$ -> \text{\hp} + e-}\ . 
\par
In our hydrodynamic simulations, we have adapted the Simplified Non-Equilibrium Cooling (SNeq) module from PLUTO \citep{tecsileanu2008simulating} in the optically thin limit to track the fraction of neutrals and ions of atomic hydrogen in the chemical reaction network equation.
In that module, PLUTO solves the equation for the neutral fraction evolution:
\begin{equation}
\label{eq:4}
\frac{ \partial x_{\text{\hi}}}{ \partial t} = n_e\left[  -(c_r+c_i)x_{\text{\hi}} +c_r\right] - J x_{\text{\hi}}\, ,
\end{equation}
where $c_r = 2.6\cdot10^{-11}\times T^{-0.5}$ and $c_i = 5.83\cdot10^{-11}\sqrt{T} \exp(-157890/T)$ are the recombination and ionisation rate coefficients in cm$^3$/s, both dependent on the temperature $T$ [K] of the gas, and on the electron number density
$n_e=(\rho/m_p) x_e$, where $m_p$ is the proton mass [g/particle]. In this work, we added the extra term in Eq. \ref{eq:4} to account for the photoionisation of the gas by the stellar XUV photons impacting the planetary atmosphere.
The formulation of the photoionisation rate coefficient $J$ [s$^{-1}$] is given in \S \ref{sec:HeatingAndCooling}. 
\par
Our model considers a single thermal temperature to describe the kinetic energy of the neutral, ion and electron components of the gas. This tacitly assumes that the transfer of kinetic energy between them occurs very rapidly, which should be a valid approximation for the relevant fractional ionisations.
\par
\subsection{Radiative transfer}
\label{sec:RadiativeTransfer}

\begin{figure}[h]
\centerline{\includegraphics[width=\linewidth]{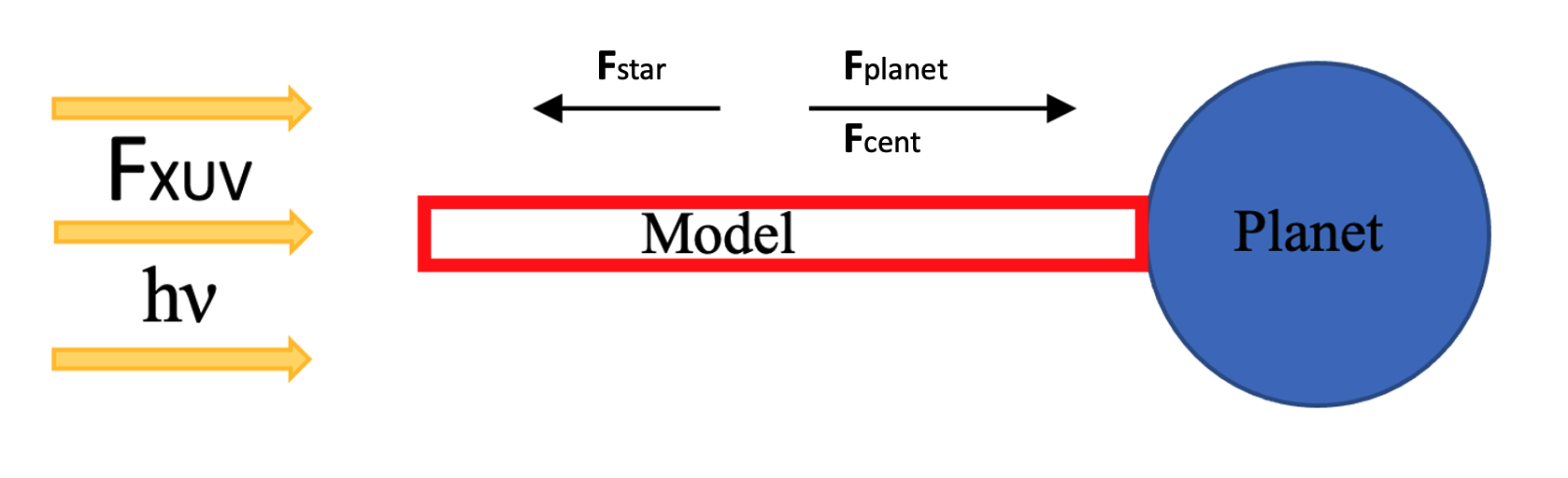}}
\caption{Illustration of the model of photoionisation and heating of an atmosphere. The atmosphere is irradiated by XUV stellar photons with an incident flux Fxuv. The gas particles are under the influence of stellar gravity ${\bf F}_{\rm star}$, planet gravity ${\bf F}_{\rm planet}$ and centrifugal forces ${\bf F}_{\rm cent}$.
}
\label{fig:sketch}
\end{figure}

Photoionisation is the only heating source of the hydrogen gas included in our model. It is caused by XUV photons coming from the star (see fig \ref{fig:sketch}). In principle, the XUV stellar spectrum should consider all possible wavelengths below 912 {\AA}, which is the threshold at which ground-state H atoms can absorb. In practice, at sufficiently short wavelengths the atmosphere of a planet becomes thin, and these wavelengths can safely be neglected in the radiative transfer. In this work, we consider a solar spectrum which starts at 15 {\AA} (see below), and we take that limit also as the shortest-wavelength in our implementation of other stellar fluxes and the H atom cross-sections.

In our simulations, we compute the local stellar flux $F_\star$($r,\lambda$) from a reference stellar flux spectrum at 1 AU, which is scaled by (1AU/$R_{\rm{orbit}}$)$^2$ to produce the top-of-the-atmosphere spectrum $F_\star(r=TOA,\lambda)$, and is attenuated with an optical depth $\tau_\lambda$ using Beer-Lambert's law. Namely:

\begin{equation}
\label{eq:5}
F_\star(r,\lambda)=F_\star(r=TOA,\lambda)e^{-\tau_\lambda(r)} \, .
\end{equation}

The wavelength-dependent optical depth $\tau_\lambda(r)$ in the direction towards the star is calculated through:

\begin{equation}
\tau_{\lambda}(r) = \sigma_\lambda  \int_{r}^{\infty} n_{\text{\hi}} d{\rm }r ,
\end{equation}

where the integral is performed along the planet-star line of sight. Here, $n_{\text{\hi}}=(\rho/m_p) x_{\text{\hi}}$ is the neutral number density [cm$^{-3}$] and $\sigma_\lambda$ [cm$^2$] the wavelength-dependent photoionisation cross-section of hydrogen, given by:

\begin{equation}
    \sigma_\lambda= \sigma_0  \left(\frac{\lambda}{\lambda_0}\right)^{3} , 
\end{equation}
where $\sigma_0$ = 6.3$\times$10$^{-18}$ cm$^2$ is the cross-section at the threshold wavelength  $\lambda_0$ = 912 {\AA}.

\begin{figure}[h]
\centerline{\includegraphics[width=\linewidth]{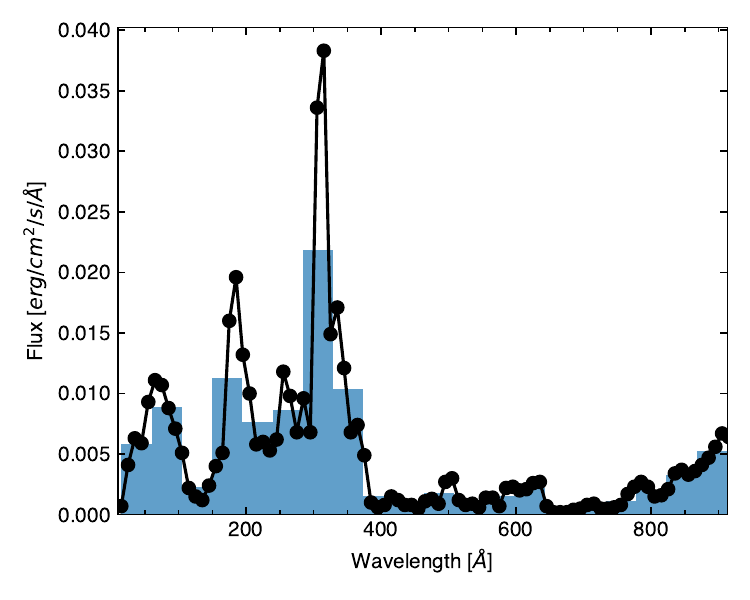}}
\caption{Solar XUV spectrum. Black: Original solar spectrum downloaded from  SOLID at 1 AU. Blue bars: Binned spectrum used in our calculations.}

\label{fig:bins}
\end{figure}

For the stellar irradiation, we adopted a solar spectrum downloaded from the SOLID\footnote{European comprehensive solar irradiance data exploitation; https://projects.pmodwrc.ch/solid} project, as observed on December 13th 2021. 
We degraded the spectrum in 20 bins of equal size in the range from $\lambda_{\rm min}=$15 {\AA} to $\lambda_{\rm 0}=$912 {\AA}. We verified that the choice of 20 bins does not impact the results presented in what follows by performing a few simulations with 40 equal-size bins, which did not show any significant difference. Figure \ref{fig:bins} displays the original solar spectrum in black along with the degraded spectrum, the latter represented by the blue bars. The XUV-integrated flux at 0.045 AU is 2174 erg/cm$^2$/s.

\subsection{Cooling, heating and ionisation rates}
\label{sec:HeatingAndCooling}

The cooling term $C$ [erg s$^{-1}$ cm$^{-3}$] in Eq.\ref{eq:3} represents the sum of the thermal energy of the electrons lost by collisional ionisation and  radiative recombination:

\begin{align}
C = n_e n_H [c_i 2.17\cdot 10^{-11} x_\text{\hi} + c_r 1.07\cdot 10^{-12}(1-x_\text{\hi})\frac{T}{11590}],
\label{eq:C1}
\end{align}
Here, $n_H = n_{\text{\hi}} + n_{\text{\hp}}$ is the total number of hydrogen nuclei, and $c_r$ and $c_i$ are the rate coefficients from Eq \ref{eq:4}. Our model does not include molecular chemistry and therefore it neglects for example cooling by H$_3^{+}$ emission in the infra-red.
In this expression, $2.17\cdot 10^{-11}$ erg is equivalent to the energy of 13.6 eV required to ionise the neutral atom. 
Correspondingly, $1.07\cdot 10^{-12} (T/11590)$ erg represents the  $\sim$0.67 kT that are extracted from the kinetic energy of the thermal electrons and lost to radiation if the environment is optically thin.
In future work, we intend to move away from the optically thin approximation and treat the recombination into the different bound levels separately, as well as keeping track of the radiated energy.

When the stellar radiation hits the planetary atmosphere, it ionises the gas and contributes to its heating. The photoionisation rate coefficient $J$ [s$^{-1}$] is given by:

\begin{equation}
\label{eq:J}
J = \int_{\lambda_{\rm min}}^{\lambda_0} \sigma_{\lambda} F_\star   \left(\frac{\lambda}{hc}\right) \left(1+\Phi_{\lambda,xe}\right) {\rm d}\lambda\, ,
\end{equation}
where $F_{\star}$ [erg cm$^{-2}$s$^{-1}${\AA}$^{-1}$] is the attenuated stellar flux described in Eq. \ref{eq:5}. $\Phi_{\lambda,xe}$ is the number of secondary ions created per photoionisation, to be described in section \ref{sec:secondary}. 
\par
The heating deposition rate $H$ [erg s$^{-1}$ cm$^{-3}$] in the energy equation can be expressed as: 
\begin{equation}
H = {n_\text{\hi}} \int_{}^{} \sigma_{\lambda} F_\star  \left(1-\frac{\lambda}{\lambda_0}\right) \eta_{\lambda,x_e}  {\rm d}\lambda\,
\label{eq:Hph}
\end{equation}
with $\eta_{\lambda,x_e}$ being the heating efficiency to be described in section \ref{sec:secondary}. Omitting the negative term in the parenthesis of the above equation, and assuming $\eta_{\lambda,x_e}$ = 1, would provide the incident energy that is deposited in the gas. The negative term in the parenthesis subtracts from it the energy that is expended to produce the primary photoelectrons during photoionisation. 
\par

Models that adopt a number of secondary ions $\Phi_{\lambda,x_e}=0$ and a heating efficiency $\eta_{\lambda,xe}=1$ assume that the surplus energy of the photoelectrons after photoionisation goes entirely into heating the gas. 
However, this is not generally true, since the photoelectrons also cause excitation and ionisation of the surrounding neutral atoms. Ideally, one has to determine precisely the heating efficiency by estimating which fraction of the energy goes into heating and which fraction is being used for the other processes. 
\par
Here, we aim to properly calculate the effect of photoelectrons on the chemistry and heating of the atmospheric gas, taking into account the dependence of these effects on the photoelectron energy and 
fractional ionisation. In section \ref{sec:secondary} we detail the physical processes in play and how to determine and parameterise $\eta_{\lambda,xe}$ and $\Phi_{\lambda,xe}$.

\subsection{Numerical Method and initial set-up}
\label{sec:1D}

To solve Eqs. \ref{eq:1}-\ref{eq:3} we use the Harten-Lax-Van Leer approximate Riemann Solver that solves exactly stationary contact discontinuities between the cells \citep[HLLC, see][]{toro2009hll}. We use linear reconstruction for the spatial order of integration and a third order Runge-Kutta scheme (RK3) for the time evolution. According to the stiffness of equations \ref{eq:3} (including cooling \ref{eq:C1} and heating \ref{eq:Hph}) and \ref{eq:4}, a dynamically-adaptive integration strategy is adopted \citep[see][for more details]{2008A&A...488..429T}.
\par

PLUTO uses dimensionless units rather than for example cgs units so that flow quantities can be properly scaled to “reasonable” numbers \citep{mignone2007pluto}. We work with three fundamental units : 
 $\rho_{0}$,  $L_{0}$ and $V_{0}$. The reference density at the base of the atmosphere is fixed for all planetary systems at $\rho_{0}$ = 1.326 $\times$ 10$^{-10}$ g/cm$^{3}$ corresponding to a pressure of 12 $\mu$bar (value taken from most models to define their boundary condition for a temperature of 1100 K. The reference length $L_{0}$ is equal to the planetary radius $R_{p}$ and the reference velocity $V_{0}$ is  calculated as the following $V_{0}$ = $(GM_{p}/L_{0})^{0.5}$. Their values are given in Table \ref{table:1} for the models considered in this study. The computational domain is composed of a stretched grid of 500 cells between 1 and 30 $R_{p}$ with a stretch factor of 1.017. The grid is extended below 1 $R_{p}$ by adding 10 cells of uniform size between 0.999 and 1 $R_{p}$. This extension is used by PLUTO to establish the boundary conditions at the bottom of our model.

\begin{table}[ht]
\centering
    \caption{Planet parameters and PLUTO parameters for each planetary system M0.69-M0.02 considered here.}
 \begin{tabular}{ccccc}
  \hline\hline
  Parameters&M0.69&M0.1&M0.05&M0.02\\
  \hline
  $M_{p}$ ($M_{\rm J}$) & 0.69 & 0.10 & 0.05 & 0.02\\
  $R_{p}$ ($R_{\rm J}$) & 1.32 & 0.69 &  0.55 & 0.40\\
  \hline
  $R_{\rm orbit}$ ($R_{\rm p}$) & 73 & 139 & 174 & 239\\
  $M_\star$/$M_{p}$ & 1593 & 10997 & 21995 & 52394\\
  $L_0$ (10$^{9}$ cm) & 9.24 & 4.83 & 3.85&2.80\\
  $V_0$ (10$^{6}$ cm/s) & 3.075 & 1.618 & 1.282&0.950\\
 \hline

 \end{tabular}
 \tablefoot{$R_{orbit}$ is the orbital distance of the planet calculated as $R_{orbit}$ = (0.045AU/$L_0$),  $M_\star$/$M_{p}$, $L_0$ and $V_0$ are the orbital distance, star-to-planet mass ratio, reference length and velocity, respectively.  
 } 
 \label{table:1}
\end{table}
\par

\begin{figure*}[ht]
\centerline{\includegraphics[width=\linewidth]{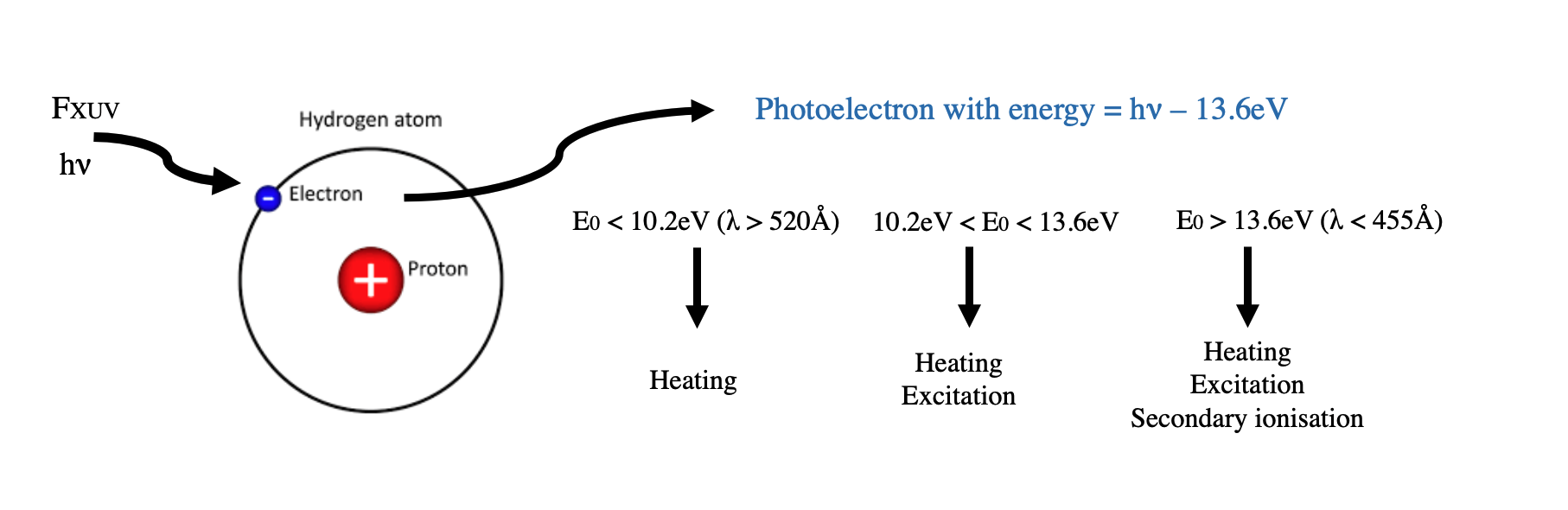}}
\caption{Schematic of the energy channelling from an impacting photon into a photoelectron that can lead to the heating of the gas, its excitation, and secondary ionisation.
}
\label{fig:scheme_photoelec}
\end{figure*}

In our simulations, the atmosphere is initialised with the density profile:
\begin{equation}
\rho(r) = \rho_{0} exp \left[\alpha_{\rho} \left(\frac{R_p}{r} - 1 \right)\right]\, ,
\end{equation}
 where $\alpha_{\rho}$ sets the initial density scale height in the atmosphere. For the simulations presented in the work, $\alpha_\rho$ is set to 20. The atmosphere is initialised everywhere at the equilibrium temperature $T_{\rm eq}$, which is equal to 1100 K in all our models in which the planet orbits a Sun-like star.
  The ideal gas law is used to initialise the pressure,
and the neutral hydrogen fraction profile is initially set as: 
\begin{equation}
x_\text{\hi} = {\rm exp} \left(\frac{R_p}{r} - 1 \right)\, ,
\end{equation}
and the velocity is set to zero.

In sections \ref{sec:outflowdescription} and \ref{sec:massdependency}, all planetary systems are considered to orbit a solar-type star with a separation of $R_{\rm orbit}$ = 0.045 AU. Table \ref{table:1} lists the physical parameters used for all simulations of each planetary systems. We choose four different planetary masses : 0.02 0.05, 0.1 and 0.69 $M_{J}$. This ranges from a sub-Neptune planet to a Jupiter-like planet similar to HD209458b. In section \ref{sec:stellarspectra}, we consider different orbital distances for the planets around different stars to ensure having the same integrated flux.
\par
At the inner boundary, we set the density and the pressure (12 $\mu$bar) to their initial value, the velocity to zero, and assume that the gas is entirely neutral. 
At the outer boundary, we use a free outflow boundary condition, with the gradients of all variables ($P$, $\rho$, $v_r$ and $x_{\rm HI}$) equal to 0.

\section{Secondary Ionisation by Photoelectrons}
\label{sec:secondary}
\subsection{Theoretical baseline}
\label{sec:theoryphotoelectrons}

Before thermalising and depositing their kinetic energy into heat, photoelectrons interact
with the neutral hydrogen atoms in the gas multiple times, losing at each collision a part of their energy. This interaction has been investigated by a number of authors for a variety of astrophysical applications
\citep{habing1971heating,shull1979heating,furlanetto2010secondary}. In the investigation of exoplanetary atmospheres, \citet{cecchi2006stellar} and \citet{shematovich2014heating} studied the effect of stellar X-ray irradiation on the heating by including the effect of photoelectrons and found that the heating efficiency is notably less than 1 if the atmosphere is mostly neutral 
when photoelectrons are included. Their simulations are not self-consistent, though, as those studies do not let the modified hydrodynamic solution alter the conditions that the photoelectrons experience. \citet{guo2016influence} investigated the production of secondary ions in H$_2$/H atmospheres, and found an increase of the total ionisation rate by a factor of 10 in the region r $<$ 1.05 $R_{p}$ that is reached only by the shortest-wavelength photons. They also report a drop in the mass loss rate by less than a factor of 2 when photoelectrons are included, although the connection with the local heating efficiency is not established.
\citet{munoz2023efficient} has looked into the photoelectron-driven processes that affect the population of the first excited level of hydrogen that is sensed in transmission spectroscopy of the {\halpha} line. 
\par
Based on the physics of the {\hi} atom excitation and ionisation, several ranges of energies may be considered. For {$E_0$}{$<$}10.2 eV, the threshold for the first excitation level, all the energy of photoelectrons is deposited as heat \citep{dalgarno1972heating,cravens1975absorption}. For energies 10.2 eV {$<$}$E_0${$<$}13.6 eV, the surplus of energy goes either into heating and into excitation. Finally, for energies $E_0$ $>$ 13.6eV,  which defines the ionisation threshold, it can additionally be used to ionise further the gas (see Fig. \ref{fig:scheme_photoelec}). The energies 10.2 eV and 13.6 eV are the thresholds for $E_0$ that enable excitation and ionisation of the atom after photoionisation, corresponding to the wavelengths 520 and 455 {\AA} respectively. For even shorter-wavelength photons, the ejected photoelectron will be able to excite/ionise the hydrogen atoms multiple times. It is reasonable to assume that the excited {\hi} (resulting from either collisions of ground state H with photoelectrons or with thermal electrons or from the recombination of protons) atom will radiate away its excitation energy. Indeed, although the Lyman-$\alpha$ line (and other lines in the Lyman series) is not optically thin, it occurs that even after accounting for opacity its effective radiative timescale is shorter than the collisional de-excitation timescale at the relevant atmospheric pressures (see \citealt{munoz2023efficient}).

\par
\begin{figure}[ht]
\centerline{\includegraphics[width=\linewidth]{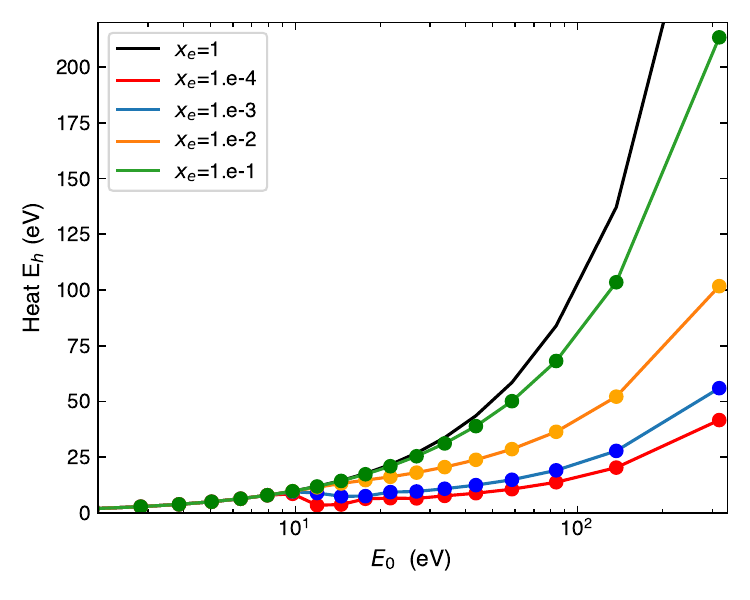}}
\caption{Energy $E_h$ deposited as heat by primary electrons of energy $E_0$. Solid curves represent different fractional ionisations. The heating efficiency is given by $\eta_{\lambda,x_e} = E_h/E_0$. Adapted from \citet{munoz2023efficient}.}
\label{fig:heateff}
\end{figure}

We calculated the heating efficiency and ionisation yield with the model described in \citet{munoz2023efficient}. The calculations were conducted over a grid in energy $E_0$ that resolves well the details in the heating efficiency and the ionisation yield at five values of the fractional ionisation, namely, $x_e$=10$^{-4}$, 10$^{-3}$, 10$^{-2}$, 10$^{-1}$ and 1. Figure \ref{fig:heateff} shows the heating $E_h$ as a function of the 
primary electron energy $E_0$ and $x_e$ (different colours). The black solid line for $x_e$ = 1 coincides with $E_h$=$E_0$. We define the heating efficiency, $\eta_{\lambda,x_e} = E_h/E_0$ which corresponds to the ratio between the energy that goes into heating $E_h$ and the initial energy of the photoelectron $E_0$. As noted earlier, for any prescribed $x_e$, the heating efficiency depends on $E_0$ or, equivalently, on the energy of the incident photon $hc/\lambda$, and takes values from 0 to 1.
\par
When $\eta_{\lambda,xe}$ = 1 all energy is deposited as heat while when $\eta_{\lambda,xe} < 1$ a fraction of the energy is diverted to excite or ionise other atoms. To facilitate its use in our hydrodynamical model and in other similar models, we fitted the MC calculations of $\eta_{\lambda,xe}$ to 4$^{\rm{th}}$ order polynomials: 
\par
\begin{equation}
\label{eq:eta_xe}
    \eta_{\lambda,xe}= \exp\left(\sum_{n=1}^{4}a_{n,\lambda}\times(\log_{10}x_e)^n\right). 
\end{equation}
The polynomial coefficients $a_{n,\lambda}$ can be found in Appendix \ref{appendix:appendix}. When the fractional ionisation falls below 10$^{-4}$, $\eta_{\lambda,x_e}$ becomes essentially independent of $x_e$. In those conditions, we fix $x_e$ to 10$^{-4}$ to estimate $\eta_{\lambda,x_e}$.

\begin{figure}[ht]
\centerline{\includegraphics[width=\linewidth]{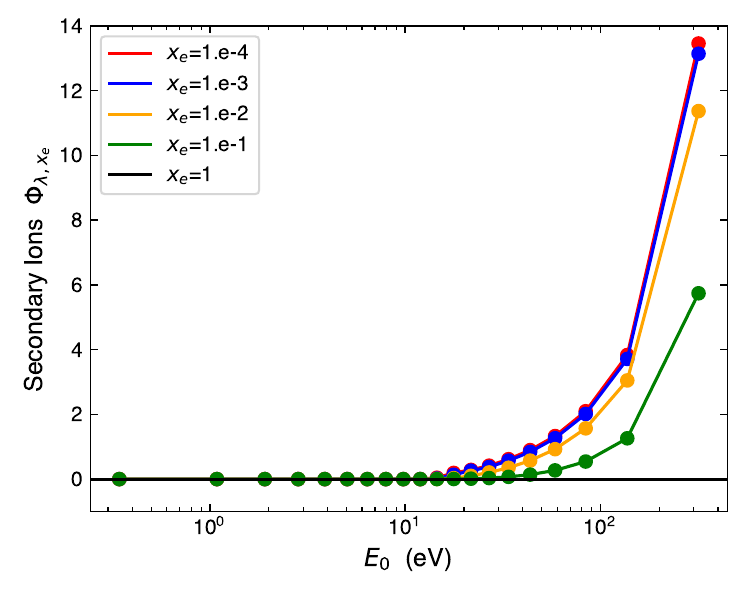}}
\caption{Secondary ions $\Phi_{\lambda,x_e}$ produced by primary electron of energy $E_0$. Solid curves represent different fractional ionisations. We note that these yields refer to energy bins of finite size rather than to the specific energies quoted there. This distinction makes some difference at the higher energies because the energy bins are larger for them. We have nevertheless confirmed by running a few tests with a refined spectral grid of 40 bins that this has no significant impact on the overall gas properties and mass loss rates. Adapted from \citet{munoz2023efficient}.
}

\label{fig:shull2}
\end{figure}

During the secondary ionisation process, a number $\Phi_{\lambda,xe}$ of secondary ions and electrons are produced from the primary electrons of energy $E_0$. Figure \ref{fig:shull2} reports Monte Carlo model calculations of the number of secondary ions created for a primary electron of energy $E_0$, and for a range of conditions of fractional ionisation $x_e$. For large photoelectron energies, numerous secondary ions can be created. For instance, for $E_0=300$ eV ($\lambda$ = 39.5 {\AA}) up to 5 secondary ions can be created if $x_e=10^{-1}$ (green curve) and up to 10 if $x_e=10^{-2}$ (orange curve). \citet{simon2011comprehensive} have shown that about 8.5 secondary ions are expected to be created at $E_0=300$ eV in a non-ionised atmosphere. Here, we consider equally-spaced bins to sample the wavelength of the impacting photons, which translates into a first wavelength bin covering 15 {\AA} (826.5 eV) to 59.85 {\AA} (207.15 eV) (see Fig. \ref{fig:bins}). In this bin, we predict the average production of about 14 secondary ions, which is significantly larger because the considered bin encompasses a range of $E_0$ that is on average larger than $300$ eV. When considering smaller wavelength bins, we recover the production of 8.5 secondary ions at $E_0=300$ eV predicted by \citet{simon2011comprehensive}. We reiterate nonetheless that our results are not significantly affected when considering smaller wavelength bins.

Conversely, for low photoelectron energies, no additional ions are created (as expected, see Fig. \ref{fig:scheme_photoelec}).
\par
Similarly, we performed a 4th order polynomial fit: 
\begin{equation}
\label{eq:phi_xe}
    \Phi_{\lambda,x_e} = \sum_{n=1}^{4}b_{n,\lambda} \times (\log_{10}x_e)^n\, ,
\end{equation}
with $b_{n,\lambda}$ the polynomial coefficients found in Appendix \ref{appendix:appendix}.

\par
Because for sufficiently large energies $E_0$, $\eta_{\lambda,xe}$ is small and $\Phi_{\lambda,xe}$ is large when $x_e$ is small, significant effects on the heating and ionisation of the atmosphere are expected in the high-pressure region where the gas remains neutral and 
that only very energetic photons can reach. Conversely, we expect that further out where the local fractional ionisation is high and the newly created photoelectrons have small energies, the effect of photoelectrons will be weak.
\par
A note is due on the terminology used throughout this paper. We will say that a calculation incorporates secondary ionisation when it is done with the parameterised forms of $\eta_{\lambda,x_e}$ and $\Phi_{\lambda,x_e}$ described above, which obviously includes the effect of both excitation and ionisation by the primary and secondary electrons. Otherwise, we will say that the calculation does not incorporate secondary ionisation when it is assumed that $\eta_{\lambda,x_e}$=1 and $\Phi_{\lambda,x_e}$=0.

\begin{figure*}[ht!]
\centerline{\includegraphics[width=0.9\linewidth]{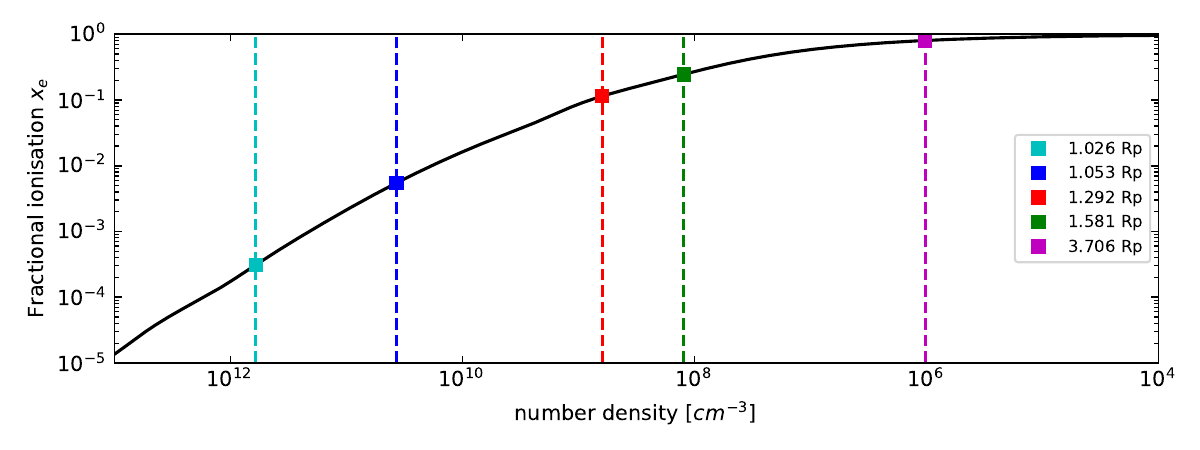}}
\centerline{\includegraphics[width=0.9\linewidth]{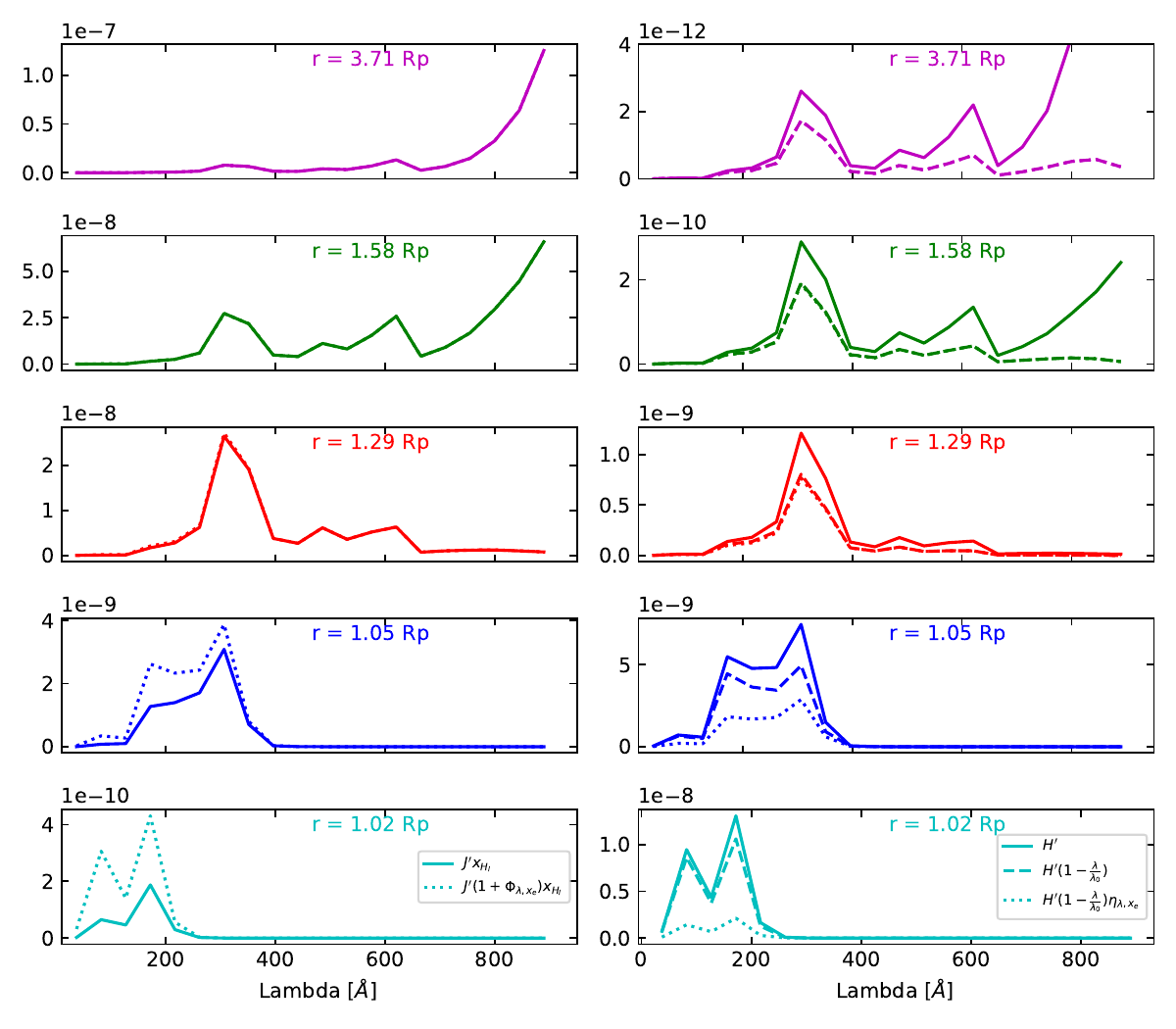}}
\caption{Expected effect of photoelectrons on ionisation and heating rates.
\textit{top panel}:
Fractional ionisation $x_e$ as a function of neutral number density $n_\text{\hi}$ in the case of a Jupiter-like planet irradiated by a solar spectrum. Five particular altitudes are labelled by coloured squares and correspond to the series of panels below.
\textit{bottom left panels}: Contribution to the photoionisation rate at different altitudes, without secondary ionisation $J$ $x_{\text{\hi}}$  [s$^{-1}$] (solid lines) and with photoelectrons (dotted lines), as a function of wavelength. The difference between the solid and dotted lines at lower altitudes clearly demonstrates the importance of multiple ionisations driven by the photoelectrons. \textit{bottom right panels}: Contribution to the heating rate at different altitudes, $H'$ [erg s$^{-1}$ cm$^{-3}$] representing the energy available before photoionisation (solid lines), the dashed line representing the heating rate due to photoionisation without the effect of photoelectrons, and the dotted lines the net heating rate taking into account secondary ionisation. Dashed and dotted lines overlap at the higher altitudes. The curves provide valuable insight into the wavelengths that contribute to heating at each altitude and how that energy is affected by photoelectrons.
}
\label{fig:J_H}
\end{figure*}

\subsection{Photoionisation and heating rates}

A main goal of this paper is to investigate the feedbacks between photoelectron-driven processes and the hydrodynamic outflow, which we detail in section \ref{sec:outflowdescription}. Before that, and to demonstrate the effects of photoelectrons, we will illustrate how they affect the energy deposition and ionisation for a prescribed atmospheric profile. For this demonstration, 
we assume the atmospheric profile to be fixed, meaning that we do not consider the feedbacks driven by the photoelectron deposition in the gas. Through this exercise, we want in particular to assess the intricate link between the shape of the solar spectrum (fig. \ref{fig:sketch}) and the deposition of the associated energy into the atmosphere. The shape of the stellar XUV spectrum partly dictates where in the atmosphere and at which wavelengths the photoionisation and heat deposition are more prominent. 
\par
For this exercise, we consider a reference profile taken from one of our numerical simulations: the steady-state atmosphere of a Jupiter-like planet with a mass of 0.69 $M_{J}$ (M0.69 in Table \ref{table:1}). The top panel of Figure \ref{fig:J_H} shows the fractional ionisation $x_e$ as a function of the neutral density $n_\text{\hi}$. In addition, we show in the bottom panels the wavelength contributions of photoionisation (left) and heating (right) at the five representative altitudes depicted in the top panel. 
\par
Each left panel of Fig.\ref{fig:J_H} shows two curves. The solid curve corresponds to $J'x_{\text{\hi}} =\sigma_{\lambda} F_\star  \left(\frac{\lambda}{hc}\right)x_{\text{\hi}}$ 
and shows the contribution to ionisation of each wavelength bin, excluding the effect of secondary ionisation by the photoelectrons. In turn, the dotted curve corresponds to $J'\left(1+\Phi_{\lambda,xe}\right)x_{\text{\hi}}$ and shows the total ionisation, including the production of secondary ions. The integral of the latter over wavelengths corresponds to the net ionisation rate $J$ $x_{\text{\hi}}$ of Eq. \ref{eq:4}.

\par
When the unattenuated stellar flux hits the upper atmosphere of the planet (see panel for 3.71 $R_{p}$, in magenta), only the longest wavelengths contribute to photoionisation, producing a large peak observed at the threshold of 912 {\AA}. This is largely because of the strong modulating effect of the photoionisation cross section, that prioritises the longer wavelengths.
As we go deeper into the atmosphere (1.58 $R_{p}$, green), the number density increases by two orders of magnitude, the atmosphere becomes optically thicker, especially at the longer wavelengths and, a broader range of wavelengths contribute to photoionisation with the presence of multiple peaks (at 300 {\AA}, 600 {\AA} and 912 {\AA}).
At 1.29 $R_{p}$ (in red), photoionisation is dominated by a peak around 300 {\AA}. The atmosphere is optically thick at the longest wavelengths, and only the ones lower than 400 {\AA} will reach this altitude, as the photoionisation cross-sections are smaller at short wavelengths. For the same reason, only the most energetic photons penetrate below 1.05 $R_{p}$ (in blue). 
Photoionisation occurs at a slower rate at this altitude than in the upper atmosphere, yet the number of electrons produced per volume unit is much larger, because it partly follows the hydrogen density.
\par
Interestingly, much faster ionisation occurs when photoelectrons are taken into account (dotted lines). The enhancement of ionisation rate by photoelectrons strengthens at the lowest altitudes. This is clearly seen at 1.05 $R_{p}$ (in blue) and 1.02 $R_{p}$ (in cyan) due to the multiplicative
term (1 +$\Phi_{\lambda,x_e}$) in the integral of $J$ $x_{\text{\hi}}$. In these deep layers, the photoelectrons enhance the double peak structure between 15 and 200 {\AA}, and ionisation is increased by a factor of two. The existence of multiple peaks results from the shape of the stellar spectrum at short wavelengths, and the extra modulating effect of $\Phi_{\lambda,x_e}$ at very short wavelengths.
\par

\begin{figure*}[ht]
\centerline{\includegraphics[width=\linewidth]{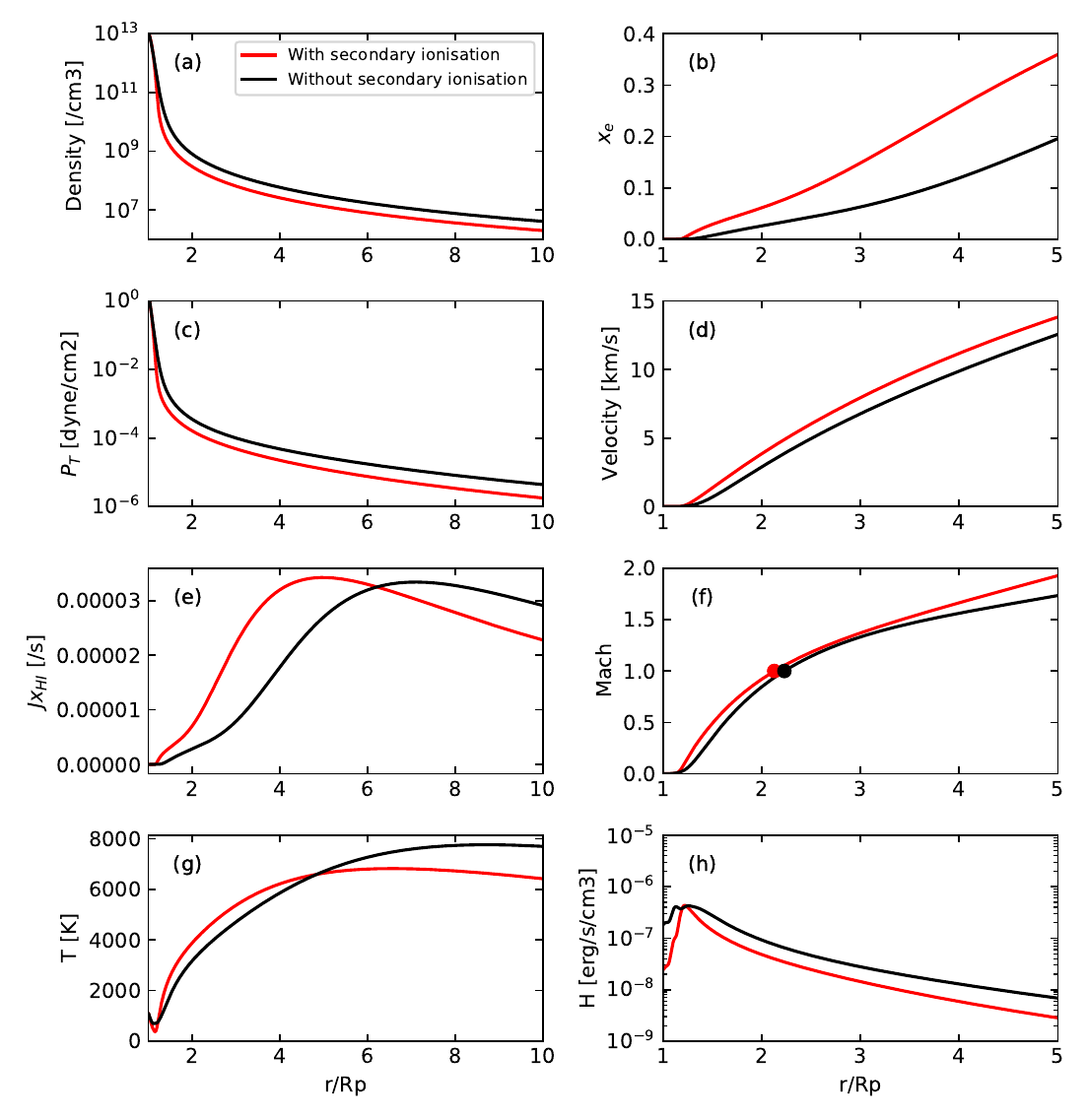}}
\caption{Evaporating atmosphere of the case M0.05 without (red lines) and with (black lines) the effect of secondary ionisation. \textit{Left panels:} from top to bottom: density, pressure, ionisation rate and temperature. \textit{Right panels:} from top to bottom: Fractional ionisation, velocity, Mach number and heating rate.}{}
\label{fig:variables}
\end{figure*}

\par
Similarly, the right panels of Fig. \ref{fig:J_H} elaborate on the energy term of Eq. \ref{eq:Hph}, and we can define three contributions. $H'={n_\text{\hi}} \sigma_{\lambda} F_\star$  depicts the energy available before photonionisation (solid lines); $H'\left(1-\frac{\lambda}{\lambda_0}\right)$ represents how much energy might potentially be effectively deposited as heat after photoionisation (dashed lines); lastly  $H'\left(1-\frac{\lambda}{\lambda_0}\right)\eta_{\lambda,x_e}$ shows the total heating rate per wavelength bin, taking into account the effect of photoelectrons (dotted lines).
\par
The energy deposition at different altitudes in the atmosphere is depicted in the right panels of fig \ref{fig:J_H}. In the upper atmosphere, for the highest altitudes, 3.71 and 1.58 $R_{p}$, the energy available before (solid line) and after (dashed line) photoionisation originates from a wide range of wavelengths. However, the heat deposition rate is 100 times larger at 1.58 $R_{p}$ than at 3.70 $R_{p}$, as a result of a higher density. For low-energy photons, the term $\left(1-\frac{\lambda}{\lambda_0}\right)$ is close to 0. Those photons spend most of their energy in the primary ionisation event, and very limited energy remains for heating (or for further ionisation or excitation). As altitude decreases, the peak of energy deposition shifts towards shorter wavelengths, and most of the net energy extracted (dashed line) from the available energy (solid lines) originates from $\lambda$ < 400 {\AA}. The peak observed at the photoionisation threshold of 912 {\AA} at the two highest altitudes disappears because most of the energy available at these wavelengths has been used for primary ionisation at higher altitudes. For the region the closest to the surface (in cyan), we observe that the energy available before $H'$ and after $H'\left(1-\frac{\lambda}{\lambda_0}\right)$ photoionisation overlap. Indeed, for wavelengths lower than 200 {\AA}, the term $\left(1-\frac{\lambda}{\lambda_0}\right)$ is close to 1 and $E_0$ is marginally impacted after removing 13.6 eV. 
\par
An additional modulation occurs when secondary ionisation is considered, described by the dotted lines in the right panels. We observe a significant attenuation of the peaks for altitudes lower than 1.29 $R_{p}$ at wavelengths lower than 400 {\AA}. For very short wavelengths, $\eta_{\lambda,x_e}$ is rather small ($\sim$0.12) and acts as a significant modulation factor, reducing substantially the heating rate arising from those wavelengths. In other words, the contribution of X-rays towards heating is very weak at all altitudes in our investigation when secondary ionisation is taken into account.
\par
In summary, we observe that taking into account photoelectrons leads to a significantly higher ionisation rate deep in the atmosphere at short wavelengths. Concomitantly, this increase implies that less energy is deposited into heat at these low altitudes, which will naturally change the steady state of the planetary atmosphere. We now turn to the investigation of possible feedbacks in the atmosphere, which we approach through numerical simulations.

\section{Secondary ionisation of an escaping atmosphere of a Neptune-size planet}
\label{sec:outflowdescription}

\subsection{Characteristics of the escaping atmosphere}

We now focus on the self-consistent simulations based on the PLUTO code, which include the hydrodynamics and photochemistry of the gas and, optionally, a parameterised description of secondary ionisation. The implementation of this new model was validated by comparing our results in the limit $\eta_{\lambda,x_e}$=1 and $\Phi_{\lambda,x_e}$=0 of no secondary ionisation with a version of the model described in \citet{garciamunoz2007physical} adapted to the physical conditions of the current simulations. The comparison between both sets of simulations proved excellent, with errors in the main properties (mass loss rates, temperature, densities, velocities) of less than a few percent. As a novel feature, we added to our current model 
the self-consistent effect of photoelectrons, as described in sections \ref{sec:modeldescription}-\ref{sec:secondary}. We focus in this section on a Neptune-mass planet (case M0.05 in Table \ref{table:1}). Figure \ref{fig:variables} shows the temperature, density, velocity, Mach number, pressure, neutral fraction profiles, along with the ionisation and heating deposition rates as a function of distance to the planet.
\par
We will first describe the case without secondary ionisation (black curves, $\Phi_{\lambda,xe}$ = 0, $\eta_{\lambda,xe}$ = 1). The profiles show the usual features that have been reported by other models of close-in exoplanets. In particular, the density (a) and pressure (c) drop rapidly from the bottom to the top of the atmosphere, while the fractional ionisation increases monotonically (b). At 5 $R_{p}$ the gas remains mostly neutral with $x_e$ = 0.2. As the flow expands, it accelerates (d), slowly at low altitudes and faster at high altitudes, and reaches the sonic point at about 2.2 $R_{p}$, marked by a black dot on the Mach profile (f). Above this altitude, the planetary wind becomes supersonic, reaching velocities of about 13 km/s at 5 $R_{p}$. The temperature profile (g) shows a more complex behaviour. Very close to the bottom of the model, the temperature decreases with altitude because little XUV radiation reaches those layers. The availability of unattenuated XUV radiation causes the rapid increase in temperature (g), ionisation (e) and heating rate (h) that occurs immediately above. Indeed, as we move high in altitude, the temperature rises up to about 7500 K at 8.5 $R_{p}$. Higher up, the temperature drops due to the expansion of the atmosphere and a lack of energy deposition as the gas becomes fully ionised and optically thin. 
\par
When secondary ionisation is taken into account (showed by the red lines on Fig \ref{fig:variables}), the density (a) and pressure 
 (c) drop faster with altitude, whereas the fractional ionisation (b) increases significantly at all altitudes. The flow velocity (d) also increases faster than in the case without secondary ionisation, yet reaches the sonic point at about the same distance of 2.1 $R_{p}$ (marked by a red dot on the Mach profile). The temperature structure of the atmosphere is also affected when secondary ionisation is considered. Indeed, the temperature (g) peaks closer to the planet at 5 $R_{p}$ and at a lower temperature around 5800K. This cooler temperature is directly associated to the fact that photoelectrons extract some of the available energy to excite and ionise further the gas instead of heating it. Higher ionisation rates (e) cause higher fractional ionisations (b). The drop in the heating rates is particularly significant deep in the atmosphere, where the gas remains mostly neutral, and the heating efficiency is very low (see below).
\par
We calculate the mass loss rate as if the atmosphere was escaping symmetrically over all directions: 
\begin{equation}
\Dot{M} = \rho v 4 \pi r^2\, .
\end{equation}
This is just a convenient assumption adopted by some 1D models that can be easily modified if instead one assumes the gas escapes only through for example the dayside, in which case the quoted mass loss rates should be divided by 4. The calculated mass loss rate for our planet M0.05 is $\Dot{M}$ = 2.89$\times$10$^{11}$ g/s without secondary ionisation and $\Dot{M}$ = 1.43$\times$10$^{11}$ g/s with secondary ionisation, decreasing the mass loss rate by about 50$\%$. This comparison shows that the drop in density when secondary ionisation is included dominates over the increased velocities that are achieved. This 50{\%} drop in the mass loss rates can be connected to the heating efficiency of the gas ($\eta_{\lambda,xe}$ in section \ref{sec:secondary}) at the altitudes where most of the stellar energy is deposited, a point that is discussed at length in what follows. A drop in the mass loss rate of that magnitude might have significant implications on atmospheric evolution studies and also on the detectability of diagnostic lines such as {\hi} {\lalpha} and {\halpha} in transmission spectroscopy, ideas that we will explore in future work.

\subsection{Heating efficiency}

\begin{figure}[ht]
\centerline{\includegraphics[width=\linewidth]{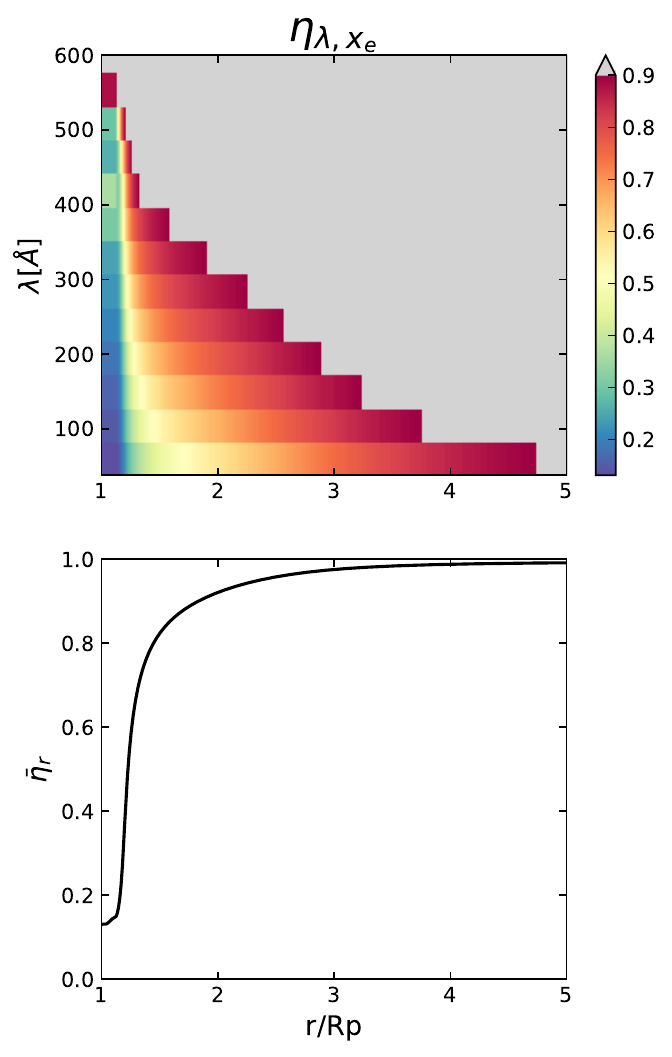}}
\caption{Heating efficiencies for the case M0.05. \textit{Top panel}: heating efficiency $\eta_{\lambda,x_e}$ as a function of wavelength and altitude. The grey regions indicate where $\eta_{\lambda,x_e}=1$. \textit{Bottom panel}: heating efficiency averaged
over wavelength as a function of altitude.}
\label{fig:eta}
\end{figure}

The amount of energy available to excite and ionise further the gas depends on the primary energy $E_0$ and on the fractional ionisation $x_e$, see section \ref{sec:secondary}. To understand better how the stellar energy is finally deposited into the atmosphere, and the wavelengths and altitudes over which this occurs, the top panel of fig. \ref{fig:eta} shows the heating efficiency $\eta_{\lambda,xe}$ for our M0.05 planet as a function of altitude and wavelength.

\par
Following fig. \ref{fig:heateff}, $\eta_{\lambda,x_e}$ = 1 (gray regions) for $\lambda$
{$>$} 520 {\AA} at every altitude due to the low energies of the primary photoelectrons, see section \ref{sec:theoryphotoelectrons}. The effect of photoelectrons is the most significant close to the surface (blue and green regions in fig \ref{fig:eta}) where the gas is largely neutral, with $x_e$ < 10$^{-3}$. $\eta_{\lambda,x_e}$ decreases steadily from 0.45 at $\lambda$ = 520 {\AA} to 0.13 for the most energetic photons ($\lambda$  < 82 {\AA}). These photoelectrons spend from 55 $\%$ to 87$\%$ of their energies to excite and ionise further the gas. We note that the heating efficiency for primary photoelectrons produced by radiation $\lambda$ < 82 {\AA} is particularly low and remains less than 0.7 up to 3.5 $R_{p}$. Higher altitudes are mildly affected by the presence of photoelectrons, and above 5 $R_{p}$ the fractional ionisation $x_e$ is very close to one, rendering the effect of photoelectrons completely negligible (grey area). Above 5 $R_{p}$ the fractional ionisation $x_e$ increases steadily and the photons ionising the atmosphere are low-energy, high wavelength photons that do not lead to significant secondary ionisation (see section \ref{sec:secondary}).
\par
The efficiency $\eta_{\lambda,x_e}$ provides insight into the specifics of wavelength-by-wavelength energy deposition in the atmosphere. To gain insight into the local heating at each altitude, even if losing some information encoded in the dependence of $\eta_{\lambda,x_e}$ on wavelength, we define in addition at each altitude an effective heating efficiency $\bar{\eta}_r$. It is calculated from the ratio of the actual heating deposition rate of Eq. \ref{eq:3} over the net energy endowed to the photoelectrons after photoionisation:

\begin{equation}
\label{eq:etaeff}
\bar{\eta}_r = \frac{\int_{}^{} \sigma_{\lambda} F_*  \left(1-\frac{\lambda}{\lambda_0}\right) \eta_{\lambda,xe}d\lambda  }{\int_{}^{} \sigma_{\lambda} F_*  \left(1-\frac{\lambda}{\lambda_0}\right) d\lambda }\, ,
\end{equation}

The bottom panel of Fig. \ref{fig:eta} shows the variation of $\bar{\eta}_r$ with altitude in the atmosphere for the M0.05 case. As altitude increases, the effective heating efficiency $\bar{\eta}_r$ increases significantly because of the increase of the fractional ionisation and because most of the nascent photoelectrons have low energies. Above 3 $R_{p}$, $\bar{\eta}_r$ is close to 1 and most of the energy is directly converted into heat without any significant effect of secondary ionisation. Below 3 $R_p$, $\bar{\eta}_r$ decreases steadily towards the bottom of the atmosphere, decreases sharply, to reach 0.5 at 1.2 $R_p$ and to reach 0.13 at the bottom.
This later latter region is particularly important to establish the mass loss rate because it is at these altitudes that the XUV photons become deposited.

\subsection{Secondary ions per primary photoelectron}

\begin{figure}[ht]
\centerline{\includegraphics[width=\linewidth]{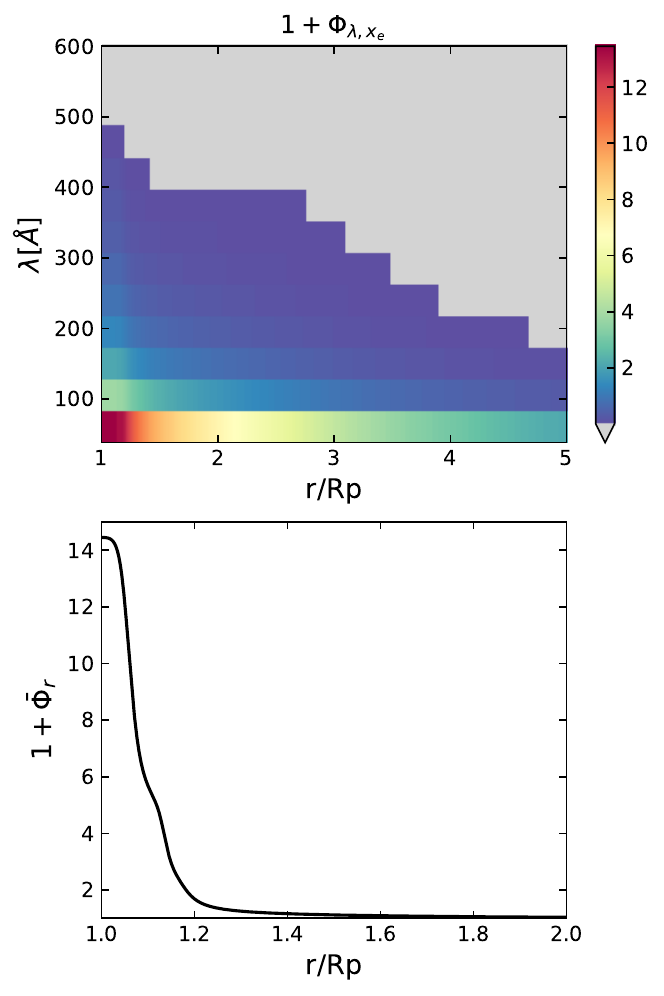}}
\caption{Ionisation yields for the case M0.05. \textit{Top panel}: number of total (primary plus secondary) ions created per primary photoelectron as a function of wavelength and altitude for our M0.05 planet. The grey regions indicate where $\Phi_{\lambda,x_e}=0$. \textit{Bottom panel}: total (primary plus secondary) ions created per primary photoelectron averaged over wavelength and as a function of altitude.
}
\label{fig:phi}
\end{figure}

Similarly to the heating efficiency, we calculate the number of total (primary + secondary) ions created per primary photoionisation event. The top panel of fig \ref{fig:phi} shows 1+$\Phi_{\lambda,xe}$ for the M0.05 planet as a function of altitude and wavelength.

\par
As anticipated, for wavelengths longer than 455 {\AA}, the primary photoelectrons will not produce any secondary ions (grey area in the top panel, see section \ref{sec:theoryphotoelectrons}). Each primary photoelectron is able to create, on average, one secondary ion for wavelengths between 455 and 306 {\AA} close to the planet when the gas is mostly neutral. For larger energies of the primary photoelectrons, 
the region where secondary ions are created extends toward higher altitude. Correspondingly, below 2 $R_{p}$, only short-wavelength photons reach those altitudes where the fractional ionisation is low, and photoelectrons will be able to create numerous secondary ions after primary ionisation. Indeed, for the shortest wavelengths below 82 {\AA}, up to 14 ions can be created per primary ionisation, reducing substantially the heating efficiency and boosting the photoionisation rate. 
\par
In order to evaluate the average number of ions created per primary energy $E_0$, an effective 1+$\bar{\Phi}_r$ is defined at each altitude. It is calculated as: 

\begin{equation}
\label{eq:phieff}
1+\bar{\Phi}_r= \frac{\int_{}^{} \sigma_{\lambda} F_*   \left(\frac{\lambda}{hc}\right) \left(1+\Phi_{\lambda,xe}\right) d\lambda }{\int_{}^{} \sigma_{\lambda} F_* \left(\frac{\lambda}{hc}\right) d\lambda }\, ,
\end{equation}

The bottom panel of fig \ref{fig:phi} shows 1+$\bar{\Phi}_r$ for the M0.05 case. The number of ions created by the primary photoelectron is close to one (i.e. no secondary photoelectrons are created) above 1.2 $R_{p}$, as expected because at those altitudes only low-energy photons interact with the gas and the fractional ionisation can be relatively high. The most striking feature of this panel is that close to the surface, up to 14 ions are created per primary ionisation. The significantly enhanced ionisation rate in that region is the ultimate cause of the faster overall ionisation of the atmosphere when secondary ionisation is considered.
\par
In summary, it is in the lower layers of the atmosphere, where only photoelectrons of high $E_0$ are created, that the heating efficiency becomes the lowest and the number of secondary ions created becomes the highest. The gas ionises faster when the effect of photoelectrons is included, which entails that the amount of heat deposited decreases and the cross-section of the planet to XUV photon decreases as well. The diminution of heat deposition naturally translates into a more compact planetary atmosphere and a reduced net mass loss. Moreover, the decrease in the heating rate we observe is not sufficient to make non-thermal processes
dominate for planets on such close-in orbits.

\section{Dependency on planetary mass}
\label{sec:massdependency}

The survival of a planetary atmosphere depends on the incoming stellar XUV flux and how this energy is used to overcome the planet's gravitational potential. The low gravitational potential of low-mass planets makes their atmospheres more extended than higher mass planets, with a larger pressure scale-height. Depending on how the neutral density and fractional ionisation profiles change with planetary mass, the relative effect of photoelectrons can be more or less pronounced. We report in table \ref{table:2} the mass loss rate for all cases presented in this study. A reduction of 43$\%$ of the mass loss rate is observed for the case M0.69 and up to 54$\%$ in the case M0.02 and a dependency of the planetary is shown: we immediately notice that secondary ionisation is slightly more efficient for low mass planets.
\par
To investigate this aspect, we now turn to analyse the two extreme cases in our sample: a planet with $M_p=0.69 M_J$ (model M0.69) and a planet with $M_p = 0.02 M_J$ (model M0.02) which is at the mass-limit for Roche lobe overflow.
\par

\begin{figure}[ht]
\centerline{\includegraphics[width=\linewidth]{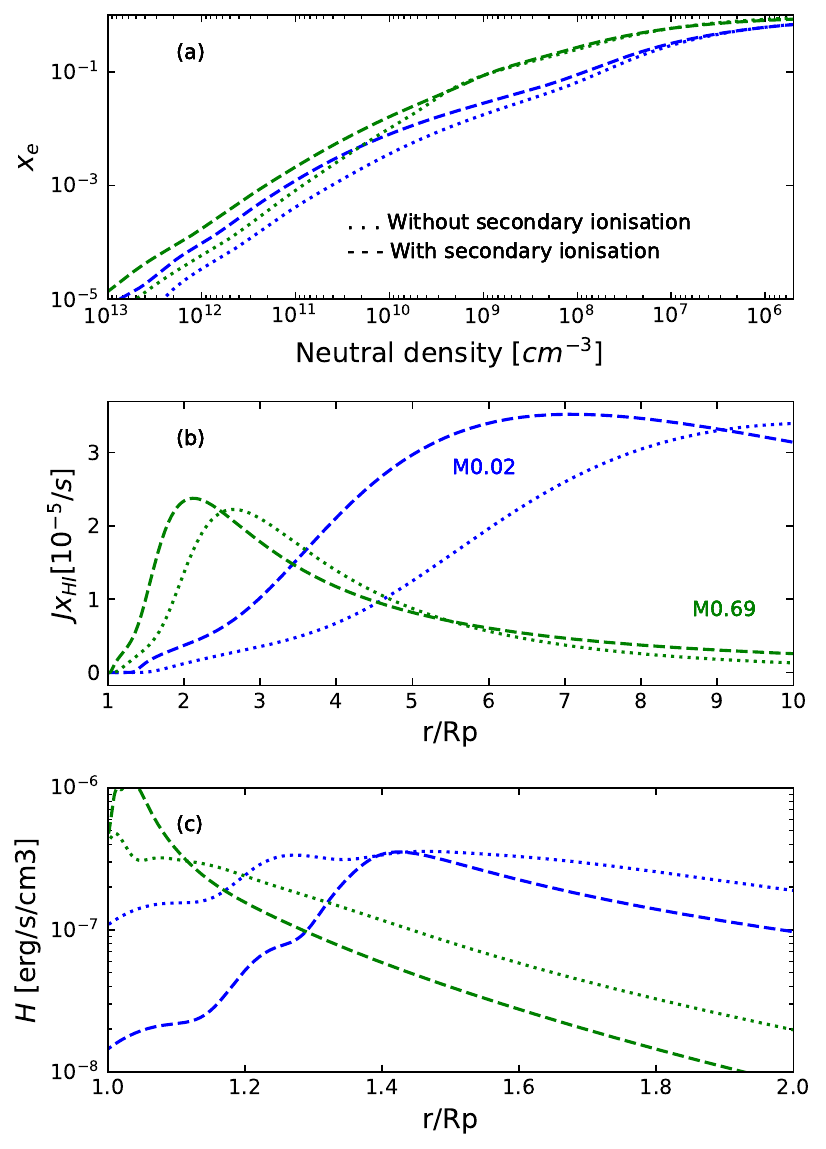}}
\caption{Comparison of two different planetary masses, with a Sub-Neptune-like planet of 0.02 $M_J$ (M0.02) and a Jupiter-like planet of 0.69 $M_J$ (M0.69). Dashed lines and dotted lines represent the cases with and without secondary ionisation, respectively. \textit{From top to bottom}: fractional ionisation $x_e$ as function of  neutral density (a), photoionisation rate $Jx_{\text{\hi}}$ (b) and heating deposition rate $H$ (c).}
\label{fig:mass}
\end{figure}
\label{sec:mass}

Both models are represented on Fig. \ref{fig:mass} with M0.69 in blue and M0.02 in green. In these figures, the dotted lines denote the cases with no secondary ionisation, and the dashed lines the case considering secondary ionisation. In panel (a) the fractional ionisation is shown as a function of neutral density. We observe that, for a given neutral density, the fractional ionisation $x_e$ is always smaller for the least massive planet (M0.02) when secondary ionisation is not included (dotted lines). As a result, one naturally expects that secondary ionisation should be more effective for the low mass planet, since its lower ionisation fraction (at a given neutral density) implies that more energy will be used to excite and ionise further the atmosphere instead of heating it. When we include secondary ionisation, this trend is indeed confirmed, and we observe a slightly larger shift of the $x_e$-$n_{\text{\hi}}$ profile for the low mass planet (M0.05, blue lines) than for the larger mass planet (M0.69, green lines).

\par
\begin{table*}[ht]
\centering
   \caption{1D spherical mass loss rates calculated for all planetary systems with and without the secondary ionisation by photoelectrons.}
\begin{tabular}{cccc}
 \hline\hline
 Planet & \multicolumn{2}{c}{$\dot{M}$ [$10^{11}$ g/s]}  & Diminution \\
  &  With secondary ionisation & Without secondary ionisation  & ($\%$)\\
 \hline
 M0.69 & 1.58 & 2.78& 43\%\\ 
 M0.1   & 1.43 & 2.75& 47\%\\
 M0.05  & 1.43 & 2.89& 50\%\\
 M0.02  & 1.51 & 3.26& 54\%\\ 
 \hline
\end{tabular}

\label{table:2}
\end{table*}

\par

In addition, we observe that in the two cases studied here, the peak of the ionisation rate (panel b) shifts to a higher altitude when secondary ionisation is included and the overall ionisation rate increases slightly. We note that these shifts and increases are more pronounced for the low-mass planet (M0.05, in blue). Parallelly, the energy deposition rate in panel (c) is also significantly impacted in both cases, similarly to the case M0.05 (section \ref{sec:outflowdescription}). Close to the surface, when photoelectrons are taken into account (dashed lines), the heating rate is reduced by one order of magnitude for M0.02 and divided by 2 for M0.69.
As a result, the effect of secondary ionisation is found to be important for all the planets studied in this work, and to be the most pronounced for the least massive planets we considered.

\section{Atmospheric escape with different stellar spectra}
\label{sec:stellarspectra}

An accurate description of stellar spectra is fundamental for modelling atmospheric escape. However, the stellar output at XUV wavelengths is quite difficult to constrain and uncertainties remain in the stellar spectra that are adopted by models. One of the difficulties arises from the fact that the high-energy radiative output of stars is variable. The main difficulty, however, is that while the X-ray spectrum is usually measurable, the EUV part is strongly attenuated by the interstellar medium (ISM). Much work has been conducted in recent years to overcome some of these difficulties, either by exploring larger sets of stars at X-ray and FUV wavelengths or by complementing the available data with elaborate modelling of stellar coronas.
We focus here on a planet similar to the case M0.69 orbiting cool, low-mass stars, and explore the impact of their XUV spectra on the photoionisation and deposition rate into the atmospheres of close-in exoplanets.

\subsection{Stellar spectral energy of K and M type stars}

As seen above, the conversion from stellar photon energy to energy powering the atmospheric escape depends, amongst other factors, on the actual energy distribution of the primary photoelectrons. It is therefore expected that the stellar spectral energy distribution (SED) may have an impact on the macroscopic properties of planetary atmospheres.  
\par

\par

\begin{figure}[ht]

 \centerline{\includegraphics[width=\linewidth]{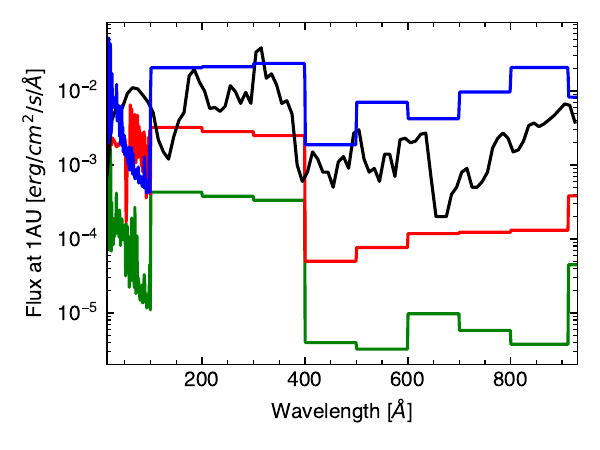}}
\centerline{\includegraphics[width=\linewidth]{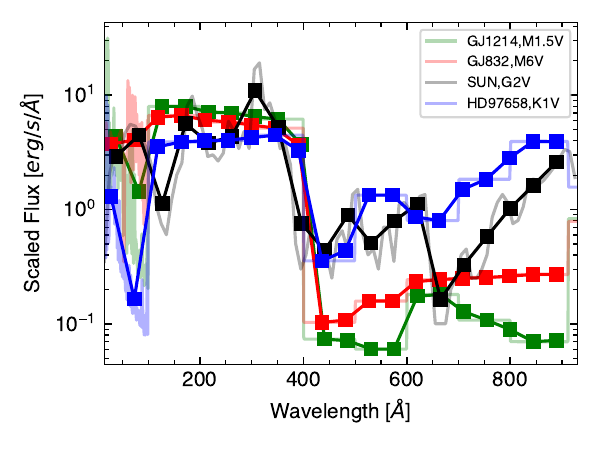}}
\caption{XUV stellar spectra. \textit{Top panel}: Panchromatic SED downloaded from the MUSCLES Treasury Survey for GJ1214, GJ832 and HD97658 and scaled at a distance of 1 AU. The solar spectrum at 1 AU used in the preceding section is shown by the black line. \textit{Bottom panel}: Scaled and binned spectra of our sample of stars. 
After scaling (see text), the integral over wavelength is the same for all four stars.}
\label{fig:stellarspectra}
\end{figure}

\begin{figure*}[ht]
\centerline{\includegraphics[width=\linewidth]{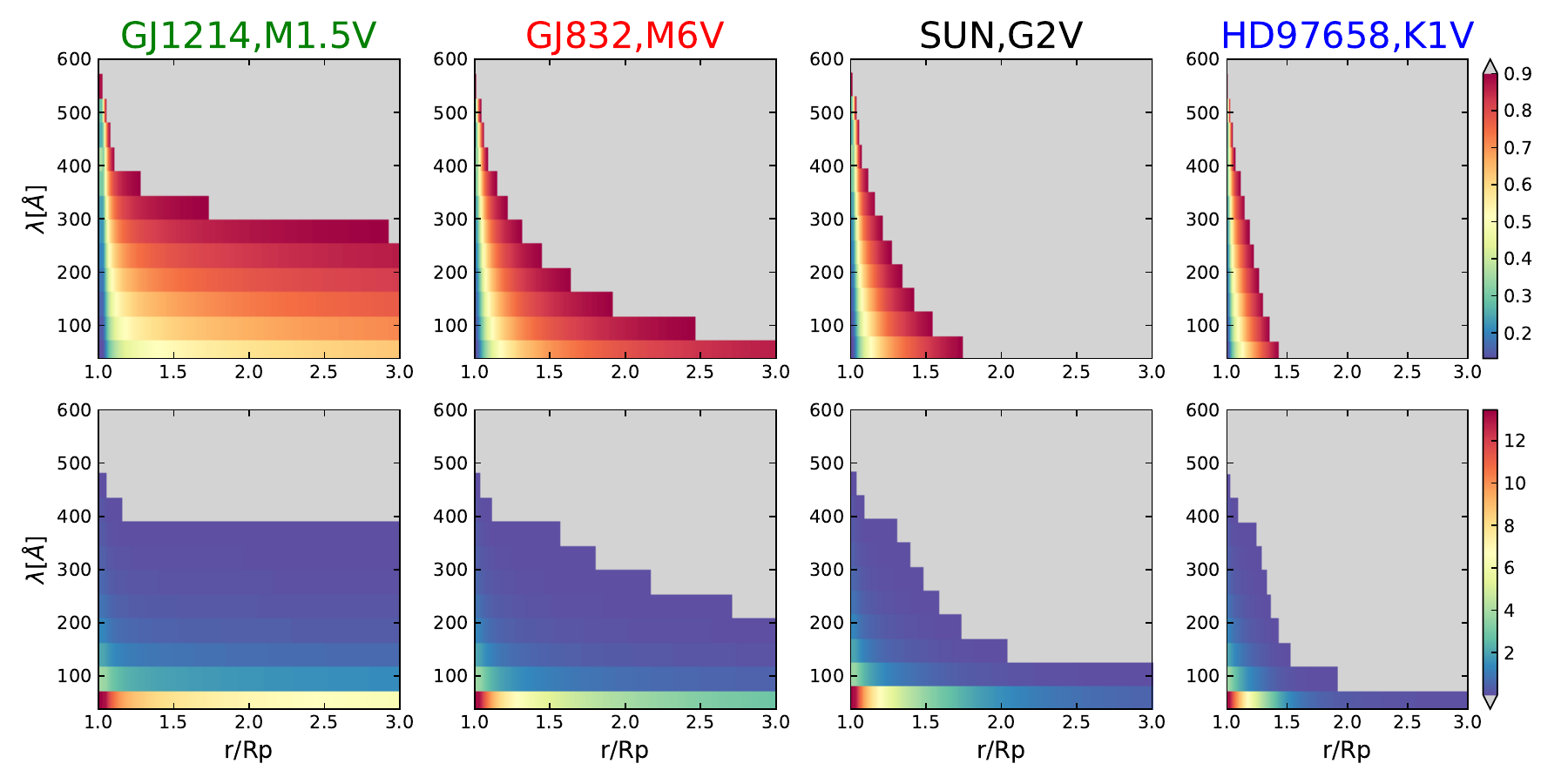}}

\caption{Heating efficiencies and ionisation yields for different stars and the Sun. \textit{Top panel}: Heating efficiencies $\eta_{\lambda,x_e}$ as function of wavelength and altitude for the four different spectra, GJ1214, GJ832, Sun and HD97658. \textit{Bottom panel}: Ionisation yields as function of wavelength and altitude.
}
\label{fig:eta_spectra}
\end{figure*}

For this purpose, we choose a sample of three stars taken from the MUSCLES data survey, GJ1214, GJ832 and HD97658 \citep{france2016muscles} (https://archive.stsci.edu/prepds/muscles/). The aim of this survey was to obtain the SEDs of M and K dwarfs from 5 {\AA} to 5.5 $\mu$m, with special emphasis on the difficult-to-constrain XUV. Typically, the X-ray part of the spectrum are measurements from Chandra and XMM-Newton \citep{loyd2016muscles}, and the EUV part was reconstructed following the techniques described by \citep{youngblood2016muscles}, who in turn used an
empirical scaling relation based on Ly$\alpha$ flux \citep{linsky2013intrinsic}. However, the X-ray spectrum for HD97658 does not come from XMM/Chandra observations but scaled from X-ray data of HD 85512 according to the ratio of bolometric flux between the two stars because of their similar levels of Fe XII emission relative to the bolometric flux. The synthetic EUV spectrum of the M dwarf GJ 832 was taken from \citet{fontenla2016semi}. We selected the panchromatic spectrum with adaptive resolution for our sample (Panchromatic SED binned to a constant 1 Å resolution, downsampled in low signal-to-noise regions to avoid negative fluxes). Choosing the spectrum without adaptive resolution does not affect our wavelength range of interest.  

\par
Similarly to the Sun, we scaled the stellar fluxes taken at 1 AU by (1AU/$R_{\rm{orbit}}$)$^2$, see section \ref{sec:RadiativeTransfer}, and degraded their spectra into 20 bins of equal sizes. In order to focus on the shape of the SED, we changed the orbital distance of the planet around each star to ensure that all planets receive the same integrated XUV flux (2174 erg/cm$^2$/s) as in the preceding sections. We display in the upper panel of Figure \ref{fig:stellarspectra} the original MUSCLES spectra, and in the lower panel the spectrum used in our model. The corresponding $R_{\rm orbit}$
are given in table \ref{table:1}. 
\par
As seen on the bottom panel of in Fig. \ref{fig:stellarspectra}, the scaled
flux of both M dwarfs is higher than the flux of the Sun and of HD97658 between 15 {\AA} and 400 {\AA}, and is lower above 400 {\AA}. Because the effect of photoelectrons is more intense at high energies (i.e. short wavelengths), we expect to have a stronger effect of the photoelectrons for the M dwarfs in our sample.
\par

\begin{figure}[ht]
\centerline{\includegraphics[width=\linewidth]{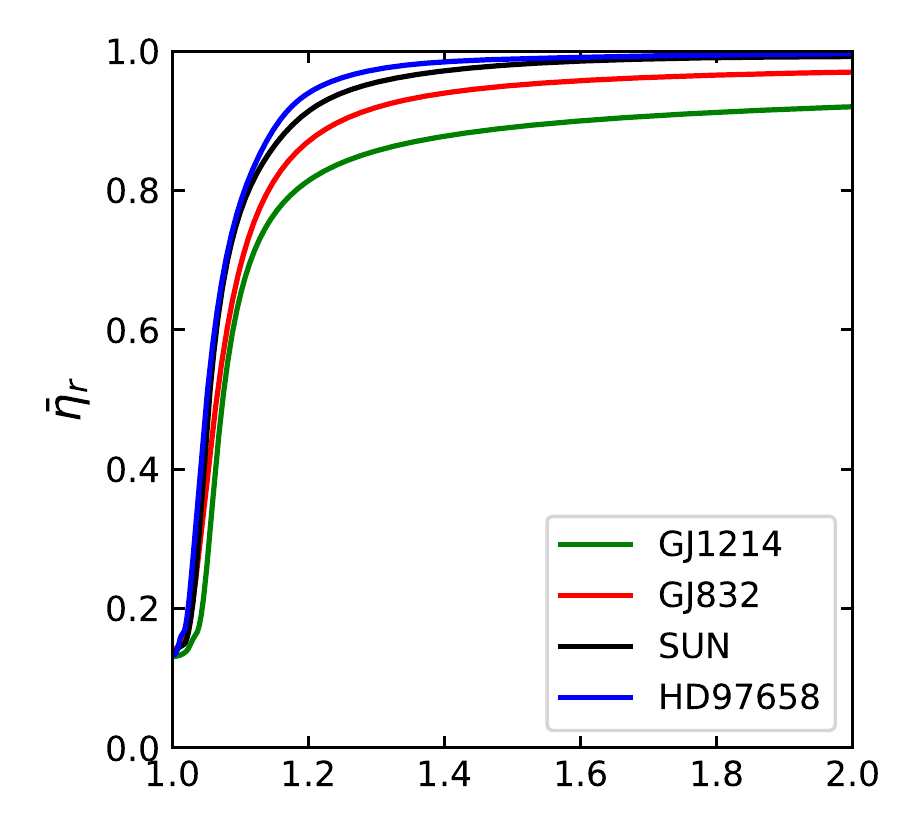}}
\caption{Heating efficiency $\bar{\eta}_r$ as a function of altitude averaged over wavelength for our sample of stars along with the Sun.
}
\label{fig:nr}
\end{figure}

\subsection{Heating efficiency and ionisation yield}

\begin{table*}[ht!]
\centering 
  \caption{Stellar parameters for HD97658, GJ832, GJ1214 and the Sun. 
  } 
 \begin{tabular}{lcccc}
 \hline\hline
 & GJ1214 & GJ832 & SUN & HD97658 \\
 \hline
 Spectral type & M1.5V & M6V & G2V & K1V \\
 $M_*$ ($M_{sun}$) & 0.15 & 0.45 & 1.00 & 0.77 \\
 XUV Flux at 1 AU [erg/cm$^2$/s] & 0.12 & 1.07 & 4.40 & 11.47 \\
 X-Ray Flux (\% XUV Flux) & 9.4 & 15.8 & 12.9 & 2.8 \\
 \hline
 $R_{\rm orbit}$ ($R_{\rm p}$) & 12 & 36 & 73 & 118 \\
 $M_\star$/$M_{p}$ & 239 & 717 & 1593 & 1227 \\
 \hline
 $\dot{M}$ with secondary ionisation [$10^{11}$ g/s] & 7.88 & 2.04 & 1.58 & 1.21 \\
 $\dot{M}$ without secondary ionisation [$10^{11}$ g/s] & 16.6 & 3.95 & 2.78 & 2.08 \\
 Mass loss diminution & 52\% & 48\% & 43\% & 41\% \\
 \hline
 \end{tabular}\\
  \tablefoot{Spectral type and $M_*$ are taken from \citep{france2016muscles}. $M_{p}$ and $R_{\rm p}$ refer to the M0.069 case. We also display the percentage of the XUV flux that goes into the X-ray flux ($\lambda <$ 100 {\AA}) and the 1D spherical mass loss rates with and without the secondary ionisation by photoelectrons.}
 \end{table*}

From our hydrodynamic calculations as described in section \ref{sec:secondary}, the heating efficiency $\eta_{\lambda,x_e}$ for each star-planet model is shown in the top panels of Fig. \ref{fig:eta_spectra} organised by increasing effect of secondary ionisation from left to right. Similarly to the solar case (third column), the effect of photoelectrons is the most significant close to the surface (blue and green regions in fig \ref{fig:eta_spectra}) where the gas is largely neutral, with $x_e$ < 10$^{-3}$. However, the effect of secondary ionisation is more prominent for the M dwarfs (GJ1214 and GJ832, first and second columns). A larger region of the atmosphere is impacted with $\eta_{\lambda,x_e}$ < 1. Interestingly, the results for HD97658 (last column) show similar heating efficiencies compared to the solar case, despite significant differences in their binned spectra above 400 {\AA}. Figure \ref{fig:nr} represents the heating efficiency $\bar{\eta}_r$ as function of altitude for all stars. We observe that GJ1214 and GJ832 have lower $\bar{\eta}_r$ for a given altitude (in units of $R_{\rm{p}}$). HD976458 displays a similar  $\bar{\eta}_r$ compared to our Sun.

\par
Along with the heating efficiencies, the bottom panels of figure \ref{fig:eta_spectra} show the ionisation yield $\Phi_{\lambda,x_e}$. For the M dwarfs, photoelectrons create, on average, at least one secondary ion onto a larger region of the atmosphere (blue areas) compared to the Sun and to HD97658. The effect of secondary ionisation is the most prominent for GJ1214 at the shortest wavelengths, up to 14 secondary ions are created below 3 $R_{p}$.

\par
We report in table \ref{table:2} the mass loss rate for all cases. A reduction of 52$\%$ of the mass loss rate is found for the case GJ1214 when secondary ionisation is taken into account. We, therefore, confirm that the effect of secondary ionisation by photoelectrons is stronger for the M dwarfs of our sample. 
\par

\section{Conclusion}

In this paper, we investigated the role of the XUV flux of a host star on planetary atmospheres, with a focus on the physics of photoelectrons. In order to understand how photoionisation and deposition of energy determine the structure of the planetary atmosphere, we have parametrised the excitation, secondary ionisation, and heating due to photoelectrons in a planetary atmosphere. We implemented this parametrisation in the PLUTO code to perform 1D modelling of escaping atmosphere of exoplanets orbiting a solar like star with a mass range between 0.02 and 0.69 $M_J$ assuming a constant planetary density. We studied the impact of secondary ionisation by photoelectrons on those processes, and on the net mass loss rate of the atmosphere. We characterised the importance of the fractional ionisation $x_e$ and the energy $E_0$ of the primary photoelectrons on the gas in the planetary atmosphere. We emphasised the importance of the shape of the XUV flux by considering different stellar spectra of K and M type stars from the MUSCLES data survey \citep{france2016muscles}.
\par
We have seen that the shape of the solar spectra in the XUV range affects significantly the photoionisation and the way the energy is deposited in the atmosphere. Photoelectrons lead to higher ionisation rates deep in the atmosphere, where the fractional ionisation of the gas is low. Consequently, the amount of energy that goes into heating is reduced when secondary ionisation is taken into account, and so does the mass loss rate of the atmosphere. Furthermore, we find that the stellar X-ray flux contributes significantly less to the heating of the atmosphere, and therefore contributes significantly less to setting the strength of the atmospheric escape.
\par
We found that the mass loss rate is decreased prominently from 43$\%$ to 54$\%$ upon consideration of secondary ionisation for planets with masses of 0.69 $M_J$ and 0.02 $M_J$. We find that the diminution of the net mass loss rate is slightly more important for planets with lower gravitational potential than the more massive ones.

\par
The shape of the stellar spectrum is a key parameter to understand the escape of planetary atmospheres. In particular, the spectral energy distribution of the XUV part of a solar-type star can vary greatly due to difference in mass, age, metallicity or magnetic activity. We implemented a sample of stellar spectra of K and M-type stars in our model and studied the effect of XUV shape on secondary ionisation and on the mass loss rate. We found that, for a Jupiter-like planet, the effect of secondary ionisation is stronger for the M dwarfs with lower heating efficiencies and higher ionisation yields compare to the Sun. The contribution of this effect is largely due to wavelengths smaller than 400 {\AA} that are comparatively stronger than the solar spectrum for the two M dwarfs studied in this work. The EUV flux between 100 and 400 {\AA}, which contributes the most to the heating, is higher for GJ1214 and GJ832, meaning that those stars would heat up efficiently the atmosphere of their planets. On the other hand, active stars in the X-ray ($\lambda < 100 {\AA}$) range will not necessarily accentuate the heating, since most of the energy of the photoelectrons will be used to re-ionise the gas. 

\par

To conclude, we stress that secondary ionisation is an important physical process and must be taken into account when modelling an escaping planetary atmosphere in 1D. In a future work, we will assess how photoelectrons impact outflow detection when considering more realistic 2D and 3D geometries. In addition, it remains unclear how secondary ionisation affects the atmospheric outflow when Joule heating is considered. Since more ions are created, it may, in some cases, lead to an increase of the Joule heating \citep{cohen2014magnetospheric} and impact the mass loss rate. The presence of a magnetic field or the consideration of a more complex chemical network may therefore also modify the mass loss rate, which we also aim to characterise in a subsequent work.

\begin{acknowledgements}
We acknowledge funding from the Programme National de Planétologie (INSU/PNP). A.S. acknowledges funding from the European Union’s Horizon-2020 research and innovation programme (grant agreement no. 776403 ExoplANETS-A) and the PLATO/CNES grant at CEA/IRFU/DAp.
\end{acknowledgements}

\bibliographystyle{aa}
\bibliography{sample631}{}

\begin{appendix}

\section{Polynomial coefficients}\label{appendix:appendix}

Polynomial coefficients for the parameterisation of the heating efficiency $\eta_{\lambda,x_e}$ (top table)
and ionisation yield $\Phi_{\lambda,x_e}$ (bottom table)
implemented in our hydrodynamical model. 
All polynomial coefficients go to zero 
at wavelengths longer than 520 {\AA} for $\eta_{\lambda,x_e}$, and 455 {\AA} for $\Phi_{\lambda,x_e}$, corresponding to the 
wavelength thresholds for excitation and ionisation by the photoelectrons (see section \ref{sec:theoryphotoelectrons}).

\begin{table}[!htbp]
\centering
 \begin{tabular}{c c c c c c} 
 \hline\hline
 \par
 $\lambda$(\AA) & $E_{0}$(eV) & $a_4$ & $a_3$ & $a_2$ & $a_1$\\ [1.5ex] 
 \hline
37.42 & 318.18&-1.35400e-02& -1.62006e-01& -5.62112e-01& -1.49574e-02\\
82.28& 137.29&-8.37464e-03 &-1.26840e-01& -5.23466e-01& -1.23433e-01\\
127.12& 84.07&-2.85578e-03& -8.40373e-02& -4.41321e-01& -1.51638e-01\\
171.98& 58.59&4.07899e-03& -2.77323e-02& -3.14691e-01& -1.28329e-01\\
216.82& 43.66&1.00390e-02& 2.35388e-02& -1.88274e-01 &-8.88287e-02\\
261.68& 33.84& 1.65556e-02& 8.03644e-02& -4.26680e-02& -2.61340e-02\\
306.52& 26.90&2.22694e-02& 1.34247e-01& 1.03973e-01& 4.64523e-02\\
351.38& 21.73&2.34987e-02& 1.52297e-01& 1.80739e-01 &8.63249e-02\\
396.23& 17.73&5.67590e-02 &4.00673e-01 &7.32612e-01& 4.08085e-01\\
441.08& 14.55&3.22789e-02& 2.61629e-01 &5.21497e-01& 3.01592e-01\\
485.93& 11.95&-1.18115e-02& -3.87313e-02 &-4.74579e-02& -1.70320e-02\\
 530.78& 9.79&-1.64632e-03& -6.83948e-03 &-1.01964e-02& -4.70321e-03\\
575.62& 7.97&0.00000e+00& 0.00000e+00& 0.00000e+00 &0.00000e+00\\
620.48& 6.41&0.00000e+00& 0.00000e+00& 0.00000e+00& 0.00000e+00\\
665.33& 5.06&0.00000e+00& 0.00000e+00 &0.00000e+00& 0.00000e+00\\
710.17& 3.88& 0.00000e+00& 0.00000e+00 &0.00000e+00& 0.00000e+00\\ 
755.03& 2.84&0.00000e+00& 0.00000e+00& 0.00000e+00& 0.00000e+00\\
799.88 &1.92 &0.00000e+00& 0.00000e+00& 0.00000e+00 &0.00000e+00\\ 
844.73& 1.10 &0.00000e+00& 0.00000e+00& 0.00000e+00& 0.00000e+00\\ 
889.58& 0.36&0.00000e+00& 0.00000e+00& 0.00000e+00 &0.00000e+00\\

 \hline
 \end{tabular}
   \caption{Calculated polynomial coefficient for the heating efficiency $\eta_{\lambda,x_e}$.}
 \label{table:5}
\end{table}

\begin{table}[!ht]
\centering
 \begin{tabular}{c c c c c c} 
 \hline\hline
 \par
 $\lambda$(\AA) & $E_{0}$(eV) & $b_4$ & $b_3$ & $b_2$ & $b_1$\\ [1.5ex] 
 \hline
 37.42 & 318.18&2.56025e-01 &2.16020e+00& 4.63252e+00& -3.01195e+00\\
 82.28& 137.29& 9.29542e-02 &8.32825e-01& 2.11135e+00 &1.10875e-01\\
 127.12& 84.07&5.35083e-02 &4.98117e-01 &1.36069e+00& 3.73183e-01\\
 171.98& 58.59& 3.10208e-02 &3.03158e-01& 8.85429e-01& 3.44192e-01\\
 216.82& 43.66&1.72333e-02& 1.80050e-01& 5.66767e-01& 2.65450e-01\\
 261.68& 33.84& 6.81667e-03& 8.65500e-02 &3.17683e-01 &1.68250e-01\\
 306.52& 26.90&9.83333e-04& 3.01000e-02& 1.53667e-01 &9.21500e-02\\
 351.38& 21.73&-8.55000e-03& -5.09167e-02& -5.55500e-02& -2.58833e-02\\
 396.23& 17.73&-7.30833e-03& -5.26500e-02& -9.49417e-02& -5.29000e-02\\
 441.08& 14.55&-5.95833e-04& -5.35833e-03& -1.08042e-02& -6.34167e-03\\
 485.93& 11.95& 0.00000e+00& 0.00000e+00& 0.00000e+00& 0.00000e+00\\
 530.78& 9.79& 0.00000e+00& 0.00000e+00& 0.00000e+00& 0.00000e+00\\
 575.62& 7.97& 0.00000e+00& 0.00000e+00& 0.00000e+00& 0.00000e+00\\
 620.48& 6.41& 0.00000e+00& 0.00000e+00& 0.00000e+00& 0.00000e+00\\
 665.33& 5.06& 0.00000e+00& 0.00000e+00& 0.00000e+00& 0.00000e+00\\
 710.17& 3.88& 0.00000e+00& 0.00000e+00& 0.00000e+00& 0.00000e+00\\
 755.03& 2.84& 0.00000e+00& 0.00000e+00& 0.00000e+00 &0.00000e+00\\
 799.88 &1.92& 0.00000e+00 &0.00000e+00& 0.00000e+00& 0.00000e+00\\
 844.73& 1.10 &0.00000e+00 &0.00000e+00& 0.00000e+00& 0.00000e+00\\
 889.58& 0.36& 0.00000e+00& 0.00000e+00& 0.00000e+00& 0.00000e+00\\
 \hline
 \end{tabular}
   \caption{Calculated polynomial coefficient for the ionisation yield $\Phi_{\lambda,x_e}$.}
 \label{table:6}
\end{table}


\end{appendix}

\end{document}